\documentclass[preprint,12pt]{elsarticle}

\usepackage{amssymb}

\usepackage{booktabs}
\usepackage{adjustbox}
\usepackage{tabularx}
\usepackage{float}
\usepackage{hyperref}

\journal{Computer Science Review}

\begin{document}

\begin{frontmatter}

\title{A review of Digital Twins and their application in Cybersecurity based on Artificial Intelligence}

\author[inst1]{MohammadHossein Homaei\corref{cor1}}
\cortext[cor1]{Corresponding author}
\ead{mhomaein@alumnos.unex.es}
\author[inst1]{\'Oscar Mogoll\'on-Guti\'errez}
\author[inst1]{Jos\'e Carlos Sancho N\'u\~{n}ez}
\author[inst1]{Mar \'Avila Vegas}
\author[inst1]{Andr\'es Caro Lindo}

\affiliation[inst1]{organization={Universidad de Extremadura},
            addressline={School of Technology, Av/ Universidad S/N}, 
            city={Caceres},
            postcode={10003}, 
            state={Extremadura},
            country={Spain}}
            
\begin{abstract}
The potential of digital twin technology is yet to be fully realised due to its diversity and untapped potential. Digital twins enable systems' analysis, design, optimisation, and evolution to be performed digitally or in conjunction with a cyber-physical approach to improve speed, accuracy, and efficiency over traditional engineering methods. Industry 4.0, factories of the future, and digital twins continue to benefit from the technology and provide enhanced efficiency within existing systems. Due to the lack of information and security standards associated with the transition to cyber digitisation, cybercriminals have been able to take advantage of the situation. Access to a digital twin of a product or service is equivalent to threatening the entire collection. There is a robust interaction between digital twins and artificial intelligence tools, which leads to strong interaction between these technologies, so it can be used to improve the cybersecurity of these digital platforms based on their integration with these technologies. This study aims to investigate the role of artificial intelligence in providing cybersecurity for digital twin versions of various industries, as well as the risks associated with these versions. In addition, this research serves as a road map for researchers and others interested in cybersecurity and digital security.
\end{abstract}


\end{frontmatter}

\section{Introduction}
The digital twin (DT) is the most superior advanced technology in industry 4.0. It is used today by several sectors and systems, including manufacturing, construction, healthcare, aerospace, transportation, smart cities, and even precision agriculture. Using DT concepts, applications have been made more efficient, effective, and reliable. DT technologies make it possible to create a virtual replica of the existing system and test, investigate and improve the activities, interactions, and results of various decisions made in real-world scenarios. Although today, most research and applications of DT are known as an emerging technology, this concept has a history of fifty years. When it was focused on conquering space, NASA's many functions of launching spaceships, rockets, and space missions were based on DT \cite{Mullet2021}. Of course, due to the hardware, software, and processing limitations, it only included summary processes and monitoring phenomena, but it can be considered the beginning of this phenomenon \cite{Huo2022}.

\begin{figure}[h!]
\centering
  \includegraphics[scale=0.2]{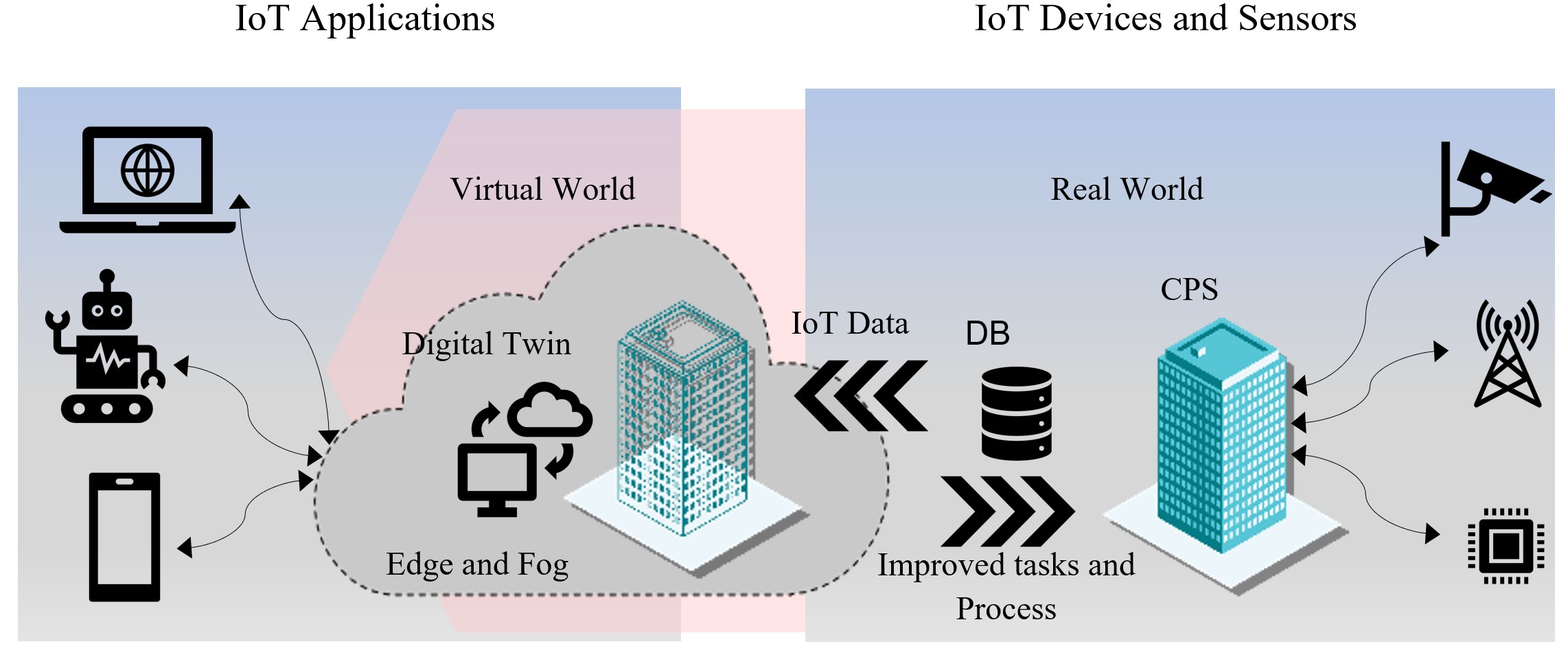}
  \caption{The relationship between DT, CPS, and IoT}
  \label{fig:1}
\end{figure}

According to IEEE author Roberto Saracco, in the last decade, the DT concept has evolved from a model of a physical entity used primarily in the design phase to a mirroring of an existing physical entity used as a reference model \cite{Saracco2022}. A physical entity's synchronisation with its associated instance may vary by leveraging embedded sensors in the physical entity, connected across a network to the DT using IoT (Internet of Things) \cite{Saracco2019}. With the increasing development of computer and communication technologies, especially IoT and Artificial Intelligence (AI) platforms, the DT concept has entered an innovation channel \cite{Homaei2022}. Perhaps in the distant years, due to the limited resources and the processing of many models, analysis and tests in the digital model were impossible or associated with very high costs, which challenged its feasibility.
DT aims to characterise physical assets through digital representation industrial and monitoring applications using detail-based techniques, mathematical models, and application programming interfaces (APIs). Almost all these things are processed and executed on servers, cloud and fog, servers, and virtual resources (such as virtual machines (VM). Digital peer processing mainly aims to predict errors and detect or apply changes and deviations. It is related to correct functioning models and things like this that often correct the natural behaviour of a model or system. Naturally, these online resources are connected to the physical world by networks such as the Internet of Things to interact with real-world components \cite{Stjepandi2021}. Therefore, in Fig. \ref{fig:1}, the connection between the physical world, cyberspace and network communication is shown.

According to the literature on DTs and their inherent difference from other classical monitoring methods, there is two-way communication between the physical and virtual worlds or a virtual model based on network communication. In classical models, the information about the physical world was sent for inspection by experts. Nevertheless, in DT models, this communication is two-way; often, the feedback is sent to the physical world or applied directly to modify the model, system, or process. According to \cite{Bergs2021}, there are three types of mirror systems, which are classified as follows:

\begin{itemize}
    \item A Digital Model (DM) is an isolated monitoring system with no automatic connection to the real world (Simulink and software model).
    \item A Digital Shadow (DS) is a system with automatic one-way communication between physical and virtual space (environmental monitoring).
     \item A DT is a system with a two-way and automatic connection between real-world and DT spaces.
\end{itemize}

An example of a DT can be seen in Fig. \ref{fig:1}. Smart buildings are characterised by physical assets (sensors, cameras, and other devices) that send information to this DT to fire the simulation model in the real world \cite{Rhm2022}. Additionally, digital assets may establish configurations and execute commands suitable for maintaining, optimising, or improving the operational performance of their physical counterparts. To achieve the above, a DT must integrate algorithms, models, patterns, equipment, and communication technologies and systems and, if necessary, make decisions that are automatically applied to physical world equipment.
Gartner also emphasises this practical aspect in its annual ranking of strategic technologies, ranking the DT paradigm fourth and first in 2018 and 2019 \cite{Saracco2022}. Market researchers predict that the DT market will grow from its current value of \$3.1 billion to \$48.2 billion by 2026. By 2031, the DT market will reach \$183 billion, with composite DTs providing the most significant opportunity \cite{Balyakin2022, Jeong2022}. According to these reports, DT applications are used in various applications, such as industrial applications, production, smart cities, agriculture and animal husbandry, disaster management, and control and monitoring of the battlefield. Therefore, the key objective of this research is to investigate the potential security threats associated with the diverse applications of DT. In classical monitoring and control systems, cybersecurity's importance often determines the application type \cite{Somers2023}. This issue is very challenging in discussing the DT because most critical infrastructure may not interest those interested in security issues at first glance. With a closer look, we find that even the most straightforward application of the DT in society's basic infrastructure can be a passive defence. Therefore, paying attention to the cybersecurity of the most basic digital applications is necessary.

\begin{figure}[h!]
\centering
  \includegraphics[scale=0.27]{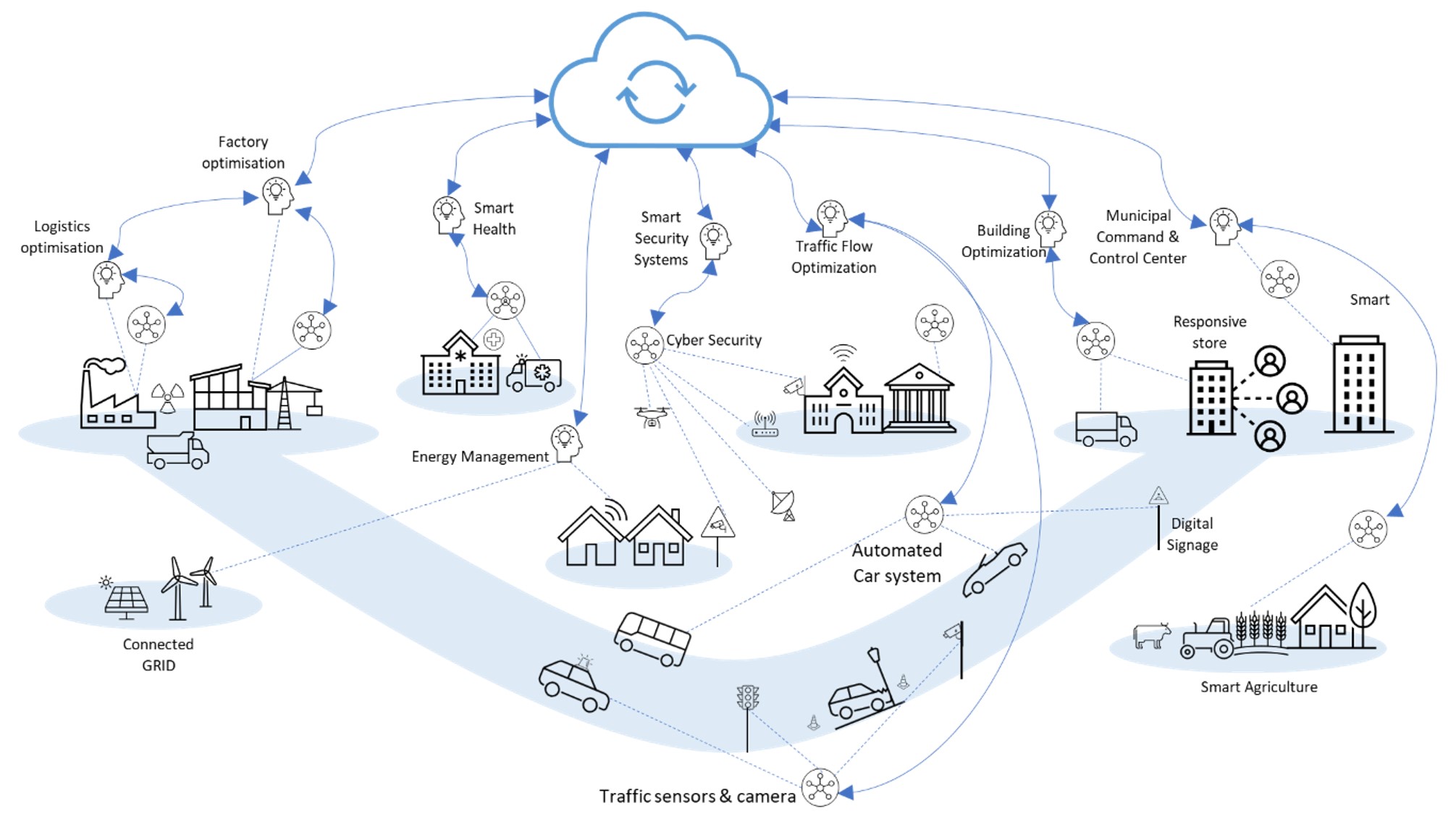}
  \caption{DT applications and Scopes (Adapted from source: Microsoft Azure Blog on Digital Twins)}
  \label{fig:2}
\end{figure}

\begin{figure}[h!]
\centering
  \includegraphics[scale=0.20]{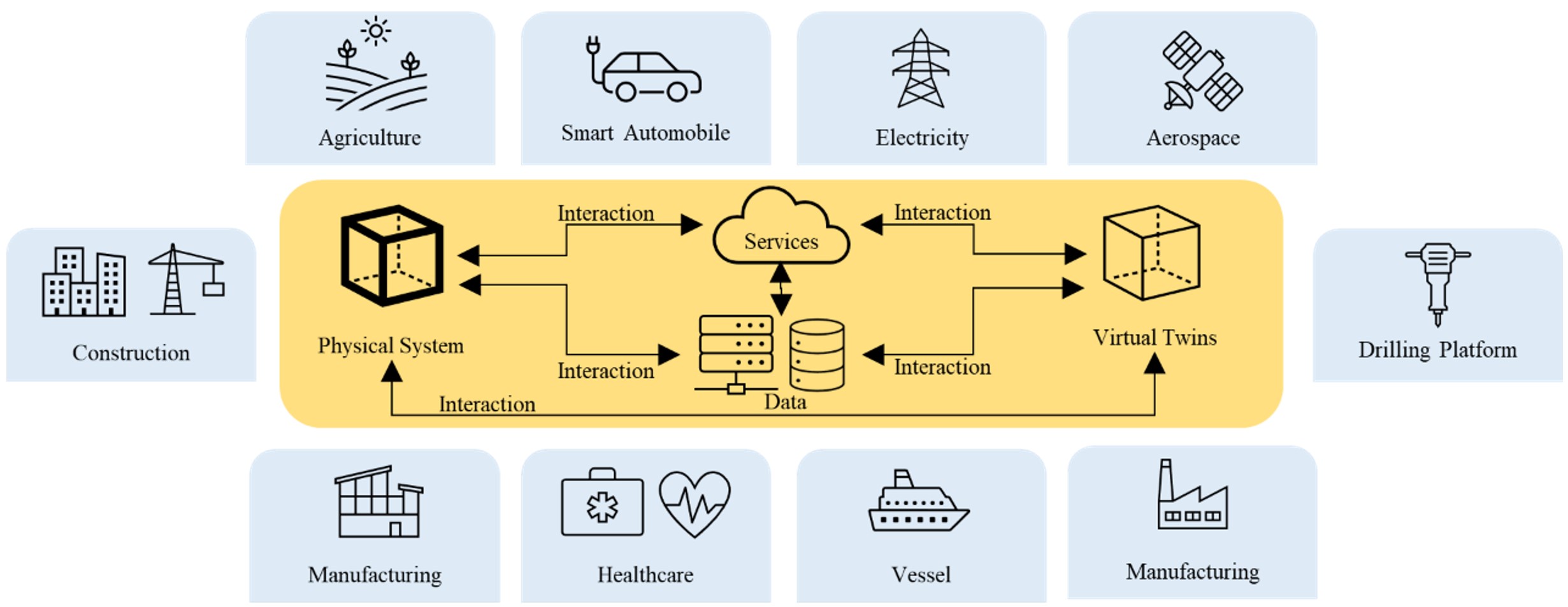}
  \caption{DT applications}
  \label{fig:3}
\end{figure}

Fig. \ref{fig:2} illustrates how each application and its tasks include essential services (data acquisition, distribution, synchronisation, modelling, simulation, representation, etcetera.) provided by various interfaces, technologies, and computation systems \cite{Korovin2022, Qian2022,Rhm2022}. Because integrating these technologies and computation systems also entails serious security risks, this paper wanted to illustrate the classification of the threats according to infrastructures of functionality and their related technologies. In most of the above applications, monitoring over the internet was once challenging. Still, today, the infrastructure is steadily growing, and some standards are being published that help stabilises communications. Industry 4.0 requires extensive cyber-physical systems (CPS) research to bridge the physical and virtual worlds \cite{Korovin2022, Shahzad2022}. This concept states that if production systems can operate more efficiently if they are intelligent to the affordability and availability of sensors and actuators, data collection has become easier than in previous decades \cite{Alcaraz2022}. However, the lack of secure platforms is also one of its challenges.
Smart cities use DTs to determine the optimal way to preserve critical assets by eliminating guesswork. DT platforms are ideal for leveraging IoT to power enterprise services and platforms. Despite its features and advantages, DT is vulnerable to cyber-attacks due to multiple attack levels and novelty, lack of standardisation and security requirements, and many reasons \cite{Guo2022A,Junior2021}. There are several cyber-attacks in the DT ecosystem, and the attack depends on the cybercriminal's goals. Our study has addressed the digital security challenges in cyberspace (Fig. \ref{fig:3}). 

There are good reasons why the cybersecurity issues of DTs have not been sufficiently explored. The most important of these reasons is that DTs are considered critical systems because they participate in automation processes, and working with them is difficult due to the issue's sensitivity. Second, DTs contain parts of intellectual property that represent a digital copy of the physical world, so most private collections in the world, to protect their business secrets, allow cybersecurity experts access to DTs of the process or production line or monitoring products or they do not provide their services \cite{Alcaraz2022}. These two aspects of subject sensitivity and copyright or protection and ownership of data and processes are desirable to cybercriminals who are trying to disrupt or harm the business of a group or organisation. Most of these applications include basic infrastructures and non-operating governmental or private defences, whose damage can endanger the security of a country.
In addition, a cybercriminal may harm DT not only from the physical environment but also from the digital space to take control of its underlying infrastructure and production assets \cite{Alshammari2021, Lv2022a}. The attack surface differs because the DT paradigm connects the two worlds through communication systems, technologies, and algorithms.
The structure of the rest of the paper is as follows: The second part identifies the basic concepts related to the idea and design of the DT, the Internet of Things, industry 4.0 and various applications of the DT. Section third includes the security challenges that are essential for the study; in the continuation of this section, the multiple requirements known in the DT cybersecurity sector are grouped in different industries because they are deeply affected by potential threats. Section four is given to the classification of solutions and AI technologies used to ensure security and increase the cybersecurity of DTs. These solutions and technologies can be part of a DT or layers of the functionality on which the DT can be based. Finally, the fifth section outlines the final remarks and future work.

\section{DT concepts and structures}
The concept of DT has become attractive to many researchers and industries; it is constantly evolving, and its definition is changing. Initially, this concept was associated with a three-dimensional simulation model of a physical object. Then, DT was considered a particular simulation based on data collected from an object or physical system. As a result, the collected data extraction was used to provide a more accurate copy at different time and space scales. At present, it can be said that DT-enabled technology refers to the ability to virtually display and optimise a physical object or a real-life system. According to \cite{Boyes2022,Fang2022,Sharma2022a,Vieira2022}, DT has three key components: physical space, cyberspace, and information processing space, which connect the two spaces above. Physical space consists of various objects (devices) or systems isolated and distributed in different regions interconnected by communication technologies. Sensors and communication equipment are used to collect and transfer multiple data types, including measurement data, user data, features/status, or device/system errors, to cyberspace. Physical space will also intelligently respond to feedback sent by cyberspace through specialised equipment such as operators. Information processing space relates physical space to cyberspace through the following processes: data storage (acquisition), data processing (pre-processing, analysis and extraction, integration), and data mapping (analysis, correlation, Sync). Cyberspace consists of virtual models (i.e., workflow model, simulation model) based on data provided by information processing space. The data of these models are used by a DT application subsystem, which is a virtual representation consisting of the visualisation and simulation of objects, or physical systems \cite{Alcaraz2022}. 

\subsection{DT Structures}
\begin{figure}[h!]
\centering
  \includegraphics[scale=0.30]{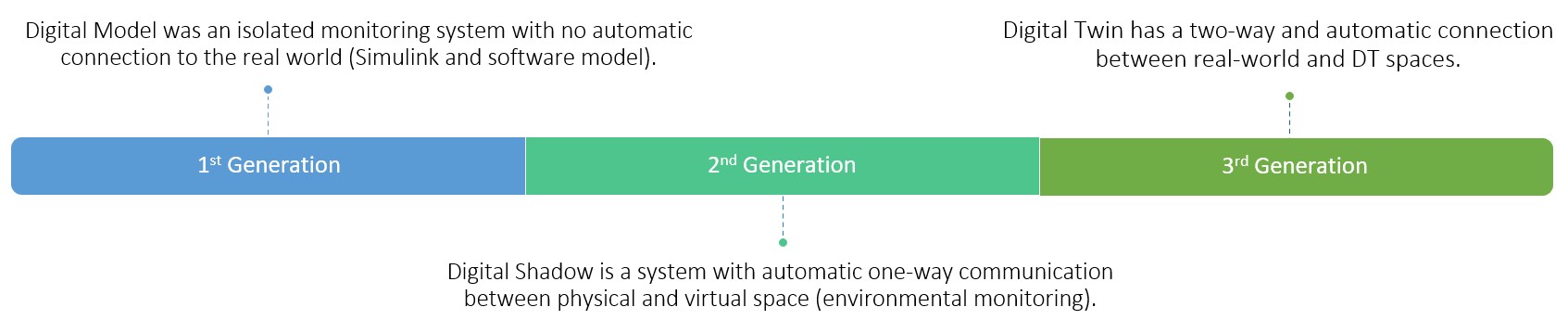}
  \caption{DT generations}
  \label{fig:4}
\end{figure}

In general, each DT consists of three components: the physical, the digital, and the connection between them. Physical systems represent any real-world scenario, such as smart grids, smart conveyance, smart industries, and cities are examples of physical systems which can provide services to multiple users. Nevertheless, when a physical system is in operation, the environment may change \cite{DavilaDelgado2021,Holmes2021,Ozkaya2022,Rudskoy2021,Stjepandi2021}. This situation requires a physical system to manage the changes, which will be very challenging, despite updating the real-world cyber-physical model based on a complex operational protocol. Thus, DT allows for the control of a simplified physical system and the simulation of new environmental data, resulting in the physical system being guided on the head due to DT. As a result of the way data is shared between physical and digital systems, DT can be categorised into three types or generations: DM, DS, and DT \cite{DavilaDelgado2021,Stjepandi2021}. Fig. \ref{fig:4} illustrates the three generations of DT.

Generally, digital modelling refers to representing one physical system or its theoretical model in digital form. The virtual digital representation of the system may contain a detailed explanation of the physical elements and components. A key difference between a DT and a cyber-physical system is that the sensor and input data create the simulation model from which the tangible assets are derived \cite{DavilaDelgado2021}. Despite this strong connection between DM and cyber-physical systems, these models cannot automatically influence the digital counterpart due to simulation results. The DS boost extends the modelling capabilities of the DM by including simulation capabilities \cite{Krckemeier2022}. The digital illustration of a physical object or a cyber-physical entity is created through a one-way process \cite{Stjepandi2021}. In a DM to interact with the environment, it is necessary to apply environmental changes and the physical world as the input to the model. The higher this interaction is, the higher the accuracy of the model and the output of the digital model. Accordingly, in a DM, the system manager can indirectly interact with the physical environment and implement the model's results on the CPS. However, this model cannot receive or resolve inconsistencies or deviations from the DM policy in real-time. In addition to the mentioned technologies, DT is a completed version of DS and DM. In DT, unlike the previous two cases, a two-way connection is established between the physical environment and DM. The input data is directly entered into the system model from sensors and input tools, and decisions are sent from the model to the system outputs \cite{Krckemeier2022}. A two-way link can influence the physical system via digital representation. DMs can include changes to the physical system, thereby providing potential results based on the system variables. Accordingly, the control system interacts with the physical system to achieve the desired outcome \cite{Stjepandi2021}]. This article also seeks to clarify some common misconceptions about DTs (as shown in Table \ref{tab:misconceptionsDT}) so that researchers can refer to the appropriate technology and then reflect on how their skills and work can contribute to future development and provide more significant insights.

\clearpage

\begin{table}[htb!]
\scriptsize
\centering
\caption{Widespread misconceptions about the DT}
\label{tab:misconceptionsDT}
\begin{tabularx}{\textwidth}{p{0.2\textwidth}X}
\toprule
\textbf{Term}                      & \textbf{Reasons and differences}                                                                                                                                                                                                                                                                                                                                                                                                                                                                                                                                            \\ 
\hline
\textbf{Device shadow}             & IoT and cloud computing platforms are expected places to find device shadow research \cite{Wang.Z.2022}. IoT device shadows are the virtual representations of physical objects; they are services for maintaining copies of information extracted from physical devices connected to the internet.                                                                                                                                                                                                                                                                                       \\ 
\hline
\textbf{Cyber twin}                & Due to the common understanding that "cyber" is another alternative term for "digital", some researchers used the terms interchangeably~ \cite{Rodrigues2021}. It is common to hear terms such as cyber-physical system, cyber-DT, and so on \cite{Adhikari2022}. For the cyber twin or cyber-physical system to be successful, a network (internet architecture) must be closely aligned with the advancements and implementation of the Internet of Everything (IoE) \cite{Dash2022}. It is standard practice to combine the network architecture of a cyber twin or cyber-physical system with a digital thread.  \\ 
\hline
\textbf{Fidelity model/Simulation} & In simulation modelling, fidelity refers to how closely a simulation model replicates the physical product it is simulating \cite{Zhang2022}. Models/simulations are often described as high/low/core/multi-fidelity, representing different levels of fidelity or considerations when building the simulation model \cite{Purcell2023}. A common feature of DTs is their real-time dynamic data exchange between a physical object and its virtual twin, as well as high or even ultra-high fidelity.                                                                                                \\ 
\hline
\textbf{Simulation}                & From the viewpoint of the virtual twin \cite{Palensky2022}, simulation refers to the critical imitating capabilities of DT technology. Simulation indicates a broader range of models since it does not consider the real-time data exchange between the physically existing object, it is a part of the DT rather than another term for it.                                                                                                                                                                                                                                          \\ 
\hline
\textbf{Digital thread}            & Digital threads provide a continuous lifetime digital/traceable record of physical products, beginning with the innovative and designing stages and continuing into the end of their existence \cite{Voth2022}. In addition, they function as enablers of interdisciplinary information exchange and play an essential role in the digitisation process \cite{Pessoa2022}.                                                                                                                                                                                                                          \\ 
\hline
\textbf{Digital Modelling}         & Modelling is one of the essential elements of a DT, but it is not an alternative term to refer to a DT. Data is exchanged bi-directionally between the physical product and its virtual twin. Still, the information is exchanged manually, which means the virtual twin represents a certain status of the physical product because of the manually controlled synthesis process \cite{Fang2022,Tao2022}.                                                                                                                                                                               \\ 
\hline
\textbf{DS}                        & In a DS, a physically existing product is represented by its virtual twin. Still, there is only a one-way data connection between the physical product and its virtual representation, which means that the virtual twins are merely digital representations of the physical entities \cite{Bergs2021,Jarosz2022}.                                                                                                                                                                                                                                                                           \\
\bottomrule
\end{tabularx}
\end{table}

The DT service categories classify the use cases identified by the literature review (Table \ref{tab:specificationsDT}).

\clearpage

\begin{table}[htb!]
\tiny
\centering
\caption{Specifications of Service categories of DTs}
\label{tab:specificationsDT}
\begin{tabularx}{\textwidth}{>{\centering}p{0.12\textwidth}p{0.25\textwidth}XXXXXX}
\toprule
Service Categories                     & Definition                                                                                                                        & \multicolumn{6}{c}{Typical   components} \\

                                       &                                                                                                                                   & Monitoring                & Simulation                & User interface            & Learning                  & Actuator                  & Analytics                  \\
                                       \midrule
Real-time monitoring                   & Monitor and log the status of a system                                                                                            & \checkmark & -                         & \checkmark & -                         & -                         &                            \\
\midrule
Energy Consumption Analysis            & Analyse and find
  ways to minimise the energy consumption of the physical system.                                                & \checkmark & \checkmark & \checkmark & -                         & -                         & \checkmark  \\
  \midrule
System Failure Analysis and Prediction & Determine if a system needs maintenance by analysing
  the received data.                                                         & \checkmark & \checkmark & \checkmark & \checkmark & -                         & \checkmark  \\
  \midrule
Optimisation/ Update                          & Achieving optimal performance for a system and
  continuously updating the parameters.                                            & \checkmark & \checkmark & \checkmark & \checkmark & \checkmark & \checkmark  \\
  \midrule
Analysis/user operation guide & analysis and provide feedback on human-made operations.                                                                            & \checkmark & -                         & \checkmark & -                         & -                         & \checkmark  \\
\midrule
Technology Integration                 & Combining
  multiple technologies under one umbrella improve operations' efficiency and
  visibility.                            & \checkmark & \checkmark & \checkmark & \checkmark & \checkmark & \checkmark  \\
  \midrule
Virtual Maintenance                    & Provide users
  with the ability to implement different maintenance strategies virtually for
  testing with minimal interference. & -                         & \checkmark & \checkmark & -                         & -                         & \checkmark  \\
\bottomrule
\end{tabularx}
\end{table}

\subsection{Classification of DTs by process}
\begin{figure}[h!]
\centering
  \includegraphics[scale=0.30]{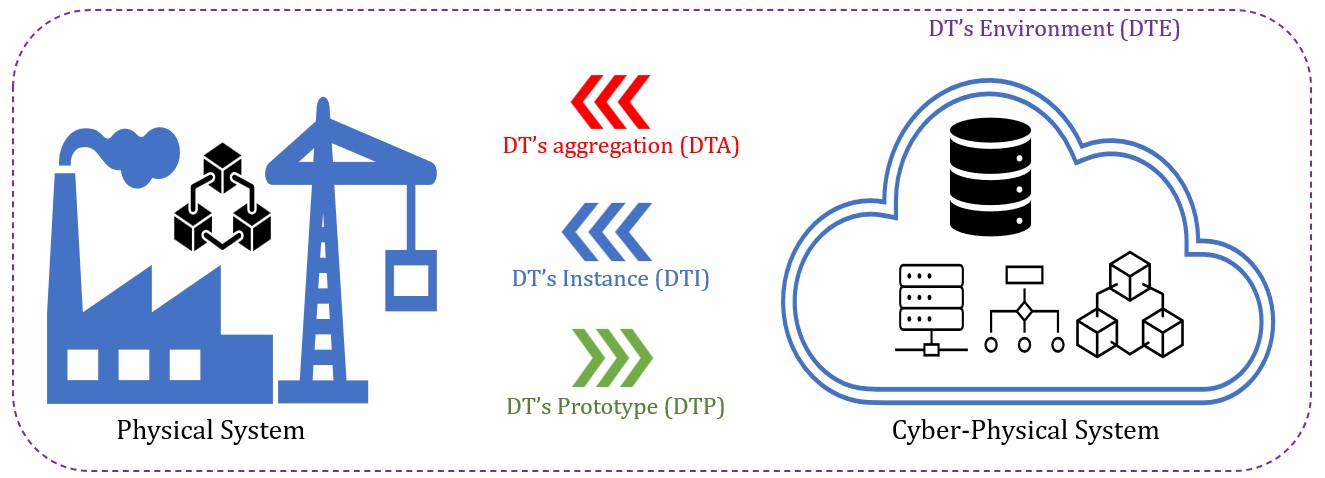}
  \caption{Four various DTs}
  \label{fig:5}
\end{figure}

It is also possible to categorise DT processes in another way, as shown in Fig. \ref{fig:5} and Table \ref{tab:typesDT}.

\clearpage

\begin{table}[htb!]
\tiny
\centering
\caption{Specifications of various types of DTs}
\label{tab:typesDT}
\begin{tabularx}{\textwidth}{p{0.1\textwidth}XXX}
\toprule
Various of DTs & Definition                                                                                                                                                                                                             & Advantages                                                                                                                                                          & Disadvantages                                                                                                                                                                                                                                          \\
\midrule
DTP            & In DTP, all CPS
  streams are used to represent the physical entities of the system. In a way,
  it is the interaction of physical assets and the virtual world.                                                       & To improve the
  efficiency of the physical system's operation by reducing costs and time. DTP
  can be used to monitor system assets in CPS.                       & DTP monitoring
  system is only focused on displaying data and processes, and it is impossible
  to change the system's efficiency.                                                                                                                    \\
  \midrule
DTI            & Unlike DTP, a DTI
  tries to communicate with CPS.DTIs handle data flow between the digital and
  physical systems, unlike DTPs. DTP and DTI establish a two-way connection
  between physical and digital techniques. & Predictions or
  guidelines are presented in the data flow, allowing the physical system to
  react simultaneously to changes in the environment as it operates.    & The disadvantage of DTI is that there is only one data
  stream from the virtual copy to the CPS.                                                                                                                                                      \\
  \midrule
DTA            & Data flow from
  the digital model to the CPS is exchanged via DTI/DTA. Based on received data
  and forecasts, CPS control can be performed.                                                                          & DTA represents the
  aggregation of all the DTIs.                                                                                                                    & Data limitations,
  hardware diversity, and infrastructural and environmental changes make it
  necessary to refine data. Therefore, several mechanisms are needed to recover
  the received data from sensors and the CPS environment for use in DT.  \\
  \midrule
DTE            & A DTE can be a
  set of DTs in an ecosystem. Each DT in DTE may have various applications.                                                                                                                             & DTE is used to
  manage complex and diverse systems in multiple DT users. The possibility of
  interaction between DTs is one of the advantages of these platforms. & Multiple digital
  systems need to be synchronised to speed up query processing.                                                                                                                                                                       \\
\bottomrule
\end{tabularx}
\end{table}

\subsection{DT and Industry 4.0}

The Industrial Revolution 4.0 revolutionised manufacturing by integrating troublesome technologies such as the IoT and cloud computing \cite{Korovin2022,RochaJcome2021,vanderBurg2021}. The increased automation and improved production synergy result from automation improved production cooperation and internet threats and attacks. Cybersecurity literature is extensive, but many actors in the manufacturing sector are just beginning to realise how crucial it is for their businesses' security \cite{Ashraf2022,Mullet2021}. Despite these security tips, a long way to go before a secure DT platform can be introduced. Most existing platforms include solutions ranging from traditional cybersecurity countermeasures to honeypot-based measures and DTs. Cyber-Physical Systems (CPS) are a typical implication of the Industrial Internet of Things (IIoT) during the I4.0 era. By combining physical equipment with multidimensional data, an intelligent awareness of the environment, and automatic control, it seeks to optimise resources and improve manufacturing efficiency. An essential characteristic of blockchain technology is its decentralisation, anti-tampering, transparency, anonymity, and contract autonomy. Due to these features, IIoT services are enhanced, and growth is promoted. In \cite{Bao2023,Wang.S.2022}, authors provide an overview of blockchain applications in the IIoT.

\subsection{DT and the IoT}
The IoT is one of the most critical parts of DT's concept. With the growing trend of technology and the increasing applications of the IoT and AI, new applications are emerging in this field every day. Due to the vastness of this network, the challenges and risks of implementing cybersecurity models in the real world are many \cite{Ashraf2022,Fortino2022,Homaei2019}. Internet-based manufacturing presents both opportunities and challenges. The information must be exchanged over a wide range of communication networks, posing questions about IT and data security that were irrelevant when the machines were not programmable and connected to any other infrastructure except the power source.
Consequently, securing the current manufacturing system in organisations is becoming more challenging due to cyberattacks and intrusions in the current environment. The security requirements for the manufacturing system are shown in the Computer-Integrated Manufacturing (CIM) model at five levels \cite{Junior2021}. CIM has been incorporated into many manufacturing models and standards as a highly integrated model.

\begin{itemize}
    \item At the enterprise/corporate level, decisions are made regarding operational management, which determines the workflows to be followed to produce the product.
    \item In the plant management network, decisions are made at the local level.
    \item Level of supervision-This level is responsible for managing various manufacturing cells, each of which performs a specific manufacturing process
    \item Process Control Level: At this level, processes perform different actions.
    \item Level of Sensor Actuators-This level consists of sensors, actuators, and controllers that work together to perform the physical process.
\end{itemize}

The design of this model makes it vulnerable to security attacks. Several protocols support this infrastructure, including Modbus, DNP3, industrial Ethernet, PROFIBUS, and BACnet, building automation and control networks. Authentication, integrity, freshness of the data, non-repudiation, confidentiality and measures to detect faults and abnormal behaviour are only used for supervisory and control mechanisms \cite{Domnguez2022}. Most manufacturers are exposed to the following cyber risks: business interruption, data breach, cyber extortion, intellectual property damage, and third-party claims. These security discrepancies can be delivered in a variety of ways, as well as:
\begin{itemize}
    \item Public Key Infrastructure - The use of device certificates and critical public infrastructure (PKI) architectures \cite{Bao2023}. Embedding PKI into embedded systems secures the communication layer, allowing the authenticity, configuration, and integrity of connected devices to be verified. Consequently, PKI is ideal for large-scale security deployments requiring a high level of security without compromising performance.
    \item By deploying anti-malware and hardening software on all IT and OT systems, highly confidential data can be encrypted to ensure that only authorised users can access it \cite{Job2022}. Moreover, symmetric encryption algorithms, hybrid encryption schemes, cryptographic hash functions, digital signatures, key agreements, and distribution protocols are widely used to ensure that only authorised entities can access the system.
    \item Monitoring the dynamic behaviour of the security systems and searching for abnormal activity is always necessary. These issues can be addressed by using intrusion detection systems (IDS). Detection techniques (the data needed for analysis) are classified as knowledge-based or behaviour-based, depending on the source of the data (the audit source). To assess the performance of an IDS, the receiver operating characteristics (ROC) curve is used, which shows the probability of detection versus the likelihood of a false alarm \cite{Kale2022}. Because of limited testbed availability and a lack of data from actual incidents, IDS research for smart manufacturing and IoT systems is in its infancy and faces many challenges.
    \item Implementing security mechanisms in smart manufacturing systems is subject to various guidelines and regulations \cite{Kaur2020}. This method includes guides such as the Guide to Industrial Control Systems (JCS) for SCADA systems, the National Institute of Standards and Technology (NIST), Distributed Control Systems (DCS), the Department of Homeland Security (DHS), the Centre for the Protection of National Infrastructure (CPNI), and so forth.
\end{itemize}

On the other hand, prominent companies in the field of information technology did not sit idle either. During the transfer of technology from physical-cyber space to digital space, they have proposed platforms that are the basis of many applications in digital cities today. Some of the most critical IoT platforms and their security challenges and solutions are listed in Table \ref{tab:DTprotocols}.

\subsection{Layers, Models, and Technologies of DT}
\begin{figure}[h!]
\centering
  \includegraphics[scale=0.33]{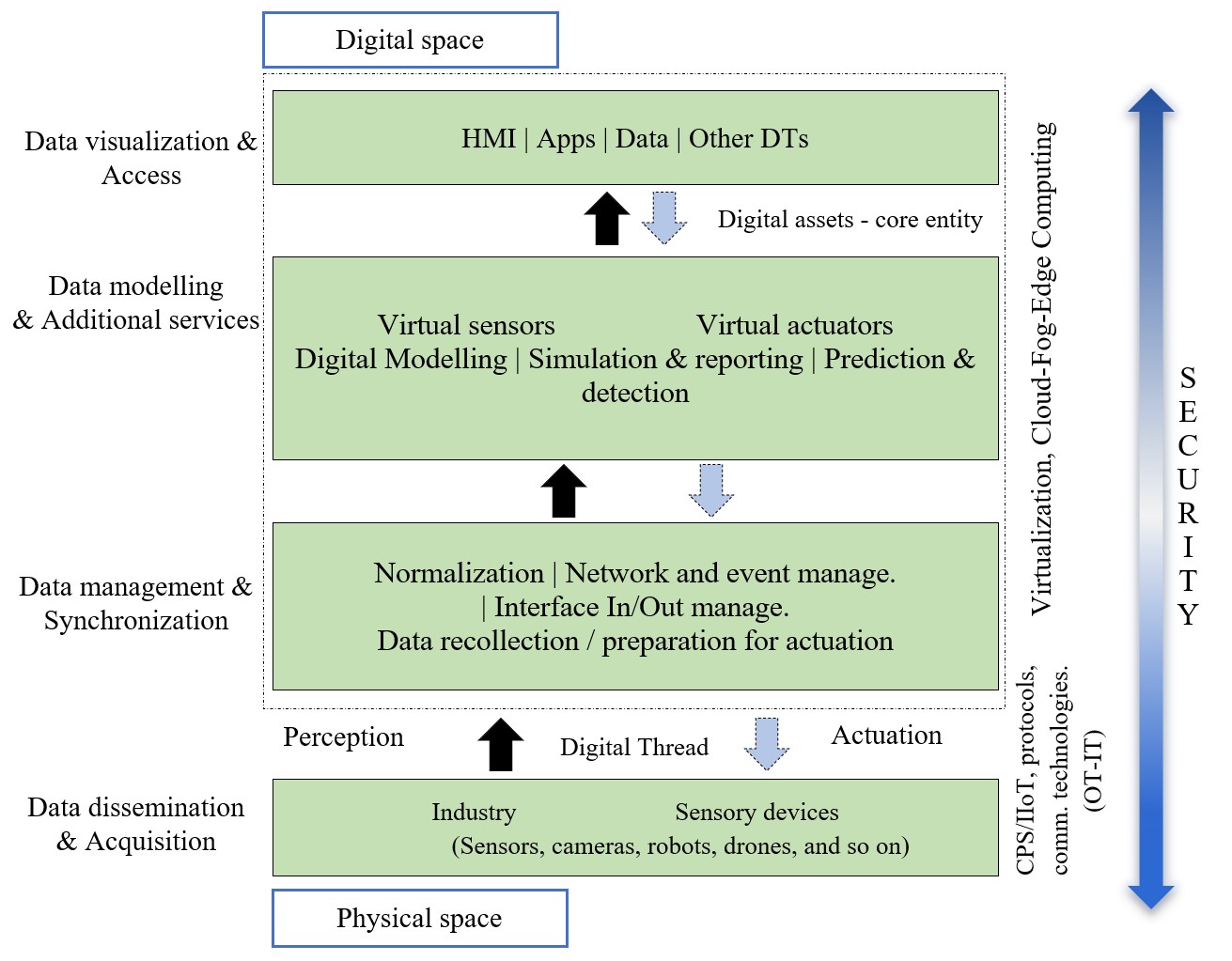}
  \caption{DTs architecture and layer (Adapted from \cite{Alcaraz2022} and modified)}
  \label{fig:6}
\end{figure}

DTs provide a means of re-creating the physical world in a digital environment, down to the tiniest details. They will be able to mimic physical objects in a digital space using advanced sensors, AI, and communication technology \cite{Homaei2020}. As shown in the previous sections, there are different approaches and little compatibility between the architecture layer models. In general, these models lack the functional granularity necessary to build a functional architecture, and some models are specifically designed for a specific purpose. The four-layer model shown in Fig. \ref{fig:6} is one of the most comprehensive layerings of DTs.

Table \ref{tab:DTprotocols} presents several standard DT protocols and their definitions, strengths, and weaknesses in the context of cybersecurity. 
\begin{table}[htb!]
\scriptsize
\centering
\caption{Popular DT protocols and their security challenges}
\label{tab:DTprotocols}
\resizebox{\linewidth}{!}{%
\begin{tabular}{>{\hspace{0pt}}m{0.1\linewidth}>{\hspace{0pt}}m{0.2\linewidth}>{\hspace{0pt}}m{0.256\linewidth}>{\hspace{0pt}}m{0.269\linewidth}>{\hspace{0pt}}m{0.302\linewidth}} 
\toprule
Protocol Name & Protocol Type        & Protocol  Characteristics                                                                                                                                                                          & Security  challenges                                                                                                                                                                                         & Solutions                                                                                                                                                                                                                                 \\ 
\midrule
CoAP  \cite{Vieira.M.N2022}        & Communication        & Explicitly developed for devices with limited resources, this protocol is based on the  User Datagram Protocol (UDP). Mechanism to facilitate the collection of continuous sensor data.          & Susceptible to  Man-in-the-Middle, Multi-vector, Unreachable destination responses,  Distributed Denial of Service (DDoS) amplified attacks, spoofing attacks,  cross-layer attacks, and reflection attacks. & CoAP does not provide authentication and authorisation; in this case, either communication security (IPsec or DTLS) or object security (within the payload) can be used.  X.509 digital certificates, pre-shared keys and public keys.  \\ 
\midrule
DTDL \cite{azure}         & Data Representation  & By depicting six characteristics of IoT elements, its open-standards platform provides seamless data transmission between various apparatuses.                                                   & Susceptible to attacks based on update distributions or modular development at the disposal of developers. (e.g., Large-scale npm  attack)                                                                  & Microsoft-managed encryption key.                                                                                                                                                                                                         \\ 
\midrule
FDTF \cite{Autiosalo2020}         & Data  Representation & The data link between DT components enables DT systems to share information using this protocol.                                                                                                 & It is not proof of the method's validity and cybersecurity side. It has many security threats, so it is under development.                                                                                 & There will be further validation of the method over time.                                                                                                                                                                                \\ 
\midrule
FIWARE  \cite{DePanfilis2018}      & Data Representation  & Various IoT components are supported for data transmission and contextual information processing.                                                                                                 & Lack of Confidentiality                                                                                                                                                                                      & -                                                                                                                                                                                                                                         \\ 
\midrule
Modbus TCP/IP \cite{Pires2021} & Communication       & Through its use of Transmission Control Protocol (TCP), this protocol enables industrial devices to be connected, provides reliable data transfer, and incorporates built-in checksum security. & Communications over networks are not confidential, there is no standard authentication, and there may be reliability problems: padding attacks, browser exploits,  protocol weakness exploits, and bugs.~  & Encapsulation of the packets, OPC UA, and IoT gateways with inbuilt Modbus/MQTT.                                                                                                                                                         \\ 
\midrule
MQTT     \cite{Human2021}     & Communication        & Internet of Things equipment can now communicate securely and reliably using this protocol based on lightweight TCP  technology.                                                                 & Authentication of data sources and anomaly detection.                                                                                                                                                        & Encryption (TLS or SSL certificates), App-Layer  Authentication (X.509), Authorisation (OAuth 2.0 (JWT), and HiveMQ).                                                                                                                     \\
\midrule
OPC UA   \cite{Abdelsattar2022}     & Data  Representation & Modelling frameworks provide information retrieval from raw data, support data manipulation, and enable monitoring.                                                                              & Message flooding, resource exhaustion, message spoofing.                                                                                                                                                    & Supported AAA  Framework and CIATriad.                                                                                                                                                                                                    \\
\midrule
URLLC    \citep{VanHuynh2022}     & Communication        & Its low latency and high reliability make it an excellent communication protocol for IoT devices.                                                                                                & Adding integrity protection may increase the computation complexity, thus adding an unacceptable delay.                                                                                                    & Using AS keys and updating the AS key based on the policy in the gNB for Intra-gNB handover.                                                                                                                                            \\
\bottomrule
\end{tabular}
}
\end{table}

\subsection{DTs Challenges}
By integrating DT into CPS, CPS can increase its efficiency through better intelligence and provide end users with a more valuable service. The DT version should accurately represent spatial entities, objects, processes, and schedules because it is supposed to be a digital copy of cyberspace. The implementation and development of such a DT present several challenges; for instance, in a network, DTs directly affect the calculation, control, and analysis steps of the output data generated by IoT equipment. Additionally, DT should be designed to meet the quality-of-service requirements, such as delay, reliability, scalability, and distribution, as well as security items, such as privacy and security. As a result of addressing these issues, new challenges and opportunities will arise concerning interdisciplinary research, which has increased its appeal \cite{BotnSanabria2022}.
\subsubsection{CPS Challenges}
CPS has strict performance requirements as a complex system requiring network, computing, and control coordination. Low latency, high reliability, and significant scalability are necessary to meet the needs of CPS applications \cite{DavilaDelgado2021,Lv2022a}. In the smart transportation system context, let us consider the development of applications related to autonomous driving. Vehicles must collect pertinent information quickly \cite{Lv2022b, Ozkaya2022}. The information and calculations received from the sensors transmitted by the communication network to and from the smart vehicle must be stored in the cloud space and the vehicle database. In addition to interacting with internal and cloud databases, the vehicle must also perform calculations rapidly. The safety of the driver, the car, and the entire transportation ecosystem depends on accurate and fast calculations in the virtual counterpart of the vehicle. In this regard, real-time service is essential, and hardware, software, and network errors must be kept at their lowest level and as close to zero as possible.
In addition, designers should ensure the scalability of the autonomous driving application they design will cope with issues such as network congestion and the complexity associated with large numbers of vehicles since the number of smart cars in each scenario will vary. It can be concluded from this example that CPS requires a high level of performance. Multiple subsystems (network, computing, and control) interact and contribute to the overall performance of the CPS \cite{Qian2022, Zhang2021}. This interaction results in a complex effect on the performance of the whole CPS system when even a slight change in the performance of one subsystem intensifies the impact on other subsystems. For CPS, it is challenging to build a DT that meets these strict performance requirements and reflects the interactions among the various subsystems.

\subsubsection{Data Science Challengess}
Mathematical modelling and data analysis are required to build a complex system's DT. The two main approaches to building an accurate DT are data-driven and model-driven. In a model-driven process, physical systems are represented using mathematical models \cite{BotnSanabria2022,Shahzad2022}. In this regard, building an accurate mathematical model of a complex physical system in CPS is challenging. The many variables in the physical system will likely complicate the mathematical and statistical model for CPS. These variables include, among others, nonlinearity, high coupling, and time variability. Using a data-driven approach, developers gather enormous amounts of information regarding the physical system's state over time. The collection and selection of these data present some challenges \cite{Qian2022}.

Ensuring that all data from different sources is equally quality is challenging. For example, in CPS, one data type may be gathered from multiple sources with varying hardware requirements. Furthermore, the storage and transmission of such a large volume of data create significant overhead for the system. The physical system's state is often sampled at a very high frequency to reflect the timely updates to the physical system. These data are estimated to require significant storage space and pose a significant performance burden on networks and computing infrastructures \cite{Alazab2022a,GrdrBroo2022,Kulik2022}. Thus, one of the fundamental problems with data science is to gather the least amount of data with the most negligible adverse effect on its effectiveness. Conclusion: The construction of a robust DT for a complex system with the smallest amount of data is a complex problem that requires both mathematical modelling and data science.

\subsubsection{Optimisation Challenges}
Undoubtedly, one of the CPS's most essential tasks is the creation of a DT that integrates computing, control, communication, and data analysis in an end-to-end chain. Especially when dealing with massively distributed networks and low-latency requirements, algorithms for allocating communication resources are incredibly complex. In addition, task offloading is another aspect of optimisation in an edge/cloud architecture \cite{Khan2022,Qian2022,Stjepandi2021}. Another optimisation problem arises due to the control mechanism, which may trigger events, time, or both. It is also necessary to extract complicated features, train them, and potentially continuously predict or classify data to meet the requirements of real-time data analysis. The DT generates a joint/integrated optimisation problem for computation, control, and communication by merging real-time computing, communication, and control processes. Consider the example of smart manufacturing supported by VR. The first step is to deliver a large amount of sensing data with the required latency. A real-time DM can then be generated using the graphics engine \cite{Guo2022A}. The control command must be executed within the latency constraint. There is a considerable challenge in optimising this VR system jointly, as any failure in one link in the chain will negatively impact the performance of the other link in the chain, even for centralised platforms with enhanced hardware.

Moreover, the distributed nature of CPS makes joint optimisation a much more difficult problem. Therefore, researchers utilised deep reinforcement learning in numerous papers to enhance the control functioning of a distributed smart manufacturing system by configuring the network and changing the sampling rate of sensors simultaneously \cite{DavilaDelgado2021,Lv2022a}. The CPS must consider more processes when DT is introduced. Two techniques are involved in the control process: the domain expert controls DT by physical objects, and the domain expert handles DT. Three computing processes are taken into account when computing DT in CPS. DT interactions are simulated in a simulation scenario, and actual CPS operations are computed based on the digital copy of a physical entity or object. DTs' communication in CPS should be considered in communication between physical objects and virtual clones, between DTs and other DTs, and between DTs and their human control interfaces. These different processes make it more challenging to solve the optimisation problem.

It is necessary to optimise the large amount of data generated during these processes to maximise the limited resources available to the CPS system. A significant concern in CPS is the data quality used to generate DTs. As a result of the need to deploy many CPS devices, the hardware requirements of one CPS device are often minimal \cite{DavilaDelgado2021,Lv2022a}. Additionally, it is possible to collect one type of data from several different sources in the CPS environment. Various CPS applications require additional data sources, which can be challenging to determine. The DT construction process often includes sampling the state of physical objects at a relatively high frequency to reflect timely updates to physical objects. The server side of DT is expected to collect unprecedented amounts of data. These data can indeed be used to assist in developing other DT systems. However, they also occupy a substantial amount of storage space. As long as the amount of stored data is not linearly related to the performance of the DT system, selecting valuable data for storage has always been challenging \cite{Holmes2021,Krckemeier2022,Ozkaya2022}. Because CPS systems are heterogeneous, IoT/CPS devices use different communication technologies. A gateway with various radio interfaces is usually deployed to exchange data between these devices. There will be additional hardware costs associated with this solution, as well as potential network bottlenecks. As the DT system is subject to strict timing and latency requirements, designing a network infrastructure to facilitate real-time communications between different components is challenging.

\subsection{Security and Privacy Challenges in DTs}
This section discusses the challenges relating to privacy and security. Since DT must constantly update the state of physical objects via network communications, it is vulnerable to cyber-attacks, which can compromise physical hardware and sensors, transmissions, and digital systems. An adversary could attack a device directly as a data collector, causing it to upload misleading or inaccurate data, an example of a data integrity attack. There is also the possibility that an adversary may compromise the gateway over which IoT machines upload/download data, intermixing invalid or misleading data with the IoT device's collected information \cite{Huo2022,Majeed2021,Qian2022}. In addition, the adversary may directly inject false data into the DT. Due to the close connection of the DT to physical objects, any attack on either will adversely affect the integrity of the whole system.

Medical records, autonomous vehicle data, and real-time smart grid operations information are all confidential information within DT applications \cite{Lalouani2022, Yu2021}. Consequently, a mechanism for authenticating cyber physical entities, digital communications, and machine-to-machine transmissions is required. Low-energy Internet of Things devices can be expensive to implement authentication mechanisms. In this situation, the authentication process has become increasingly tricky on devices with limited power. A DT application builds a digital replica of the physical system by utilising sensor information from the object layer. Attackers may attack sensors or gateways using denial-of-service (DoS) attacks or malware propagation \cite{Ghimire2022,Khraisat2021}. A compromised gateway or sensor makes it difficult for DT to obtain data regarding the overall condition of the CPS. Therefore, mitigation mechanisms must be proposed and validated to reduce the impact of such attacks.
Data from IoT sensors and gateways are required for DT applications to operate. In this case, research is needed on data and communication security, and privacy \cite{Ashraf2022,Huo2022,Wang.S.2022}. The security of the research is discussed from several perspectives, including confidentiality, integrity, and availability. Let us now consider the smart transportation system as an example of privacy. To acquire data, DT uses IoT sensors installed near roads, vehicles, and et cetera. In addition, IoT sensors may be owned and operated by various organisations and companies. The confidentiality of data must be maintained by using mixed authentication and secure communication protocols in all organisations and operators. 

Furthermore, some of the data stored at the application layer, such as the vehicle's location and the data collected by traffic cameras, are sensitive. For this reason, access control is necessary to protect users' privacy. In addition, specific security regulations may safeguard users' privacy and ensure their data's confidentiality by limiting data dissemination, especially at the end-user level. Regarding medical information sharing and privacy, the Health Insurance Portability and Accountability Act (HIPPA) will make these regulations.  
Several DT applications require real-time data to construct a DM of the system, including smart grids, smart transportation, smart manufacturing, and smart cities. Data integrity must be maintained for the DT system to model and guide the physical system effectively. Based on the DT architecture, we discuss the integrity of the four layers. Object layer communication means sensors communicating with gateways. Sensors must communicate with gateways. To prevent a man-in-the-middle attack (MITM) from being used against the sensor, the application layer communication between the sensor and gateway must be secured. Implementing a security mechanism that prevents unauthorised users from altering data is essential. Further, anomaly detection can be employed at the application layer to detect manipulated data and ensure data integrity. The final step in preventing unauthorised access to the DT is implementing an authentication system at the user level.

Furthermore, real-time DT depends on the keys and availability of sensors and gateways, which can be vulnerable to cyber-attacks such as distributed denial of service (DDoS) and malware propagation. The mitigation schemes gateways used to filter malicious traffic can reduce the impacts of attacks in DTs. Despite this, DTs may be overloaded when adversaries send time-consuming or enormous queries through the end-user layer. These threats can be mitigated by implementing mitigation schemes at the application level. Multiple users may be able to access the same service simultaneously during a specific period. Load-balancing algorithms can be incorporated into the gateway in this situation. As a result, a highly resilient system is created by using nearby low-load gateways when a gateway is overloaded.

\subsection{Security challenges by application} 
Using AI and Machine Learning (ML) is not the only challenge organisations, and cybersecurity professionals must overcome \cite{Ashraf2022,Ghimire2022,Thelen2022}. Among the weaknesses in the current approach to security are the following:
\begin{itemize}
    \item Distant infrastructure: In today's world, systems communicate across continents, sending sensitive data around the globe. It is easier to break into these transfers since they are not adequately protected.
    \item Manual detection: Security threats and suspicious patterns are not monitored 24/7 by human teams. In most cases, systems are not observed.
    \item Reactivity of security teams: Most security experts focus on confronting threats rather than predicting threats. 
    \item Dynamic treats: Hackers use many strategies to conceal their locations, IP addresses, identities, and methods. In contrast, data created by businesses are easily accessible to criminals in the cybersecurity field.
\end{itemize}

\subsubsection{Smart Cities}
With advanced technologies, the smart world will become a reality where all physical objects will be equipped with embedded computing and communication capabilities. Monitoring the process via the internet was once challenging, but today, the infrastructure is growing steadily, and some standards are releasing which help stabilise communication. Industry 4.0 will require extensive research on Cyber-Physical Systems (CPS) to bridge the physical and virtual worlds. This concept states that if production systems are smart, they can function more efficiently \cite{Guo2022A,Pessoa2022}. Data acquisition has become relatively more straightforward than in previous decades due to the affordability and availability of sensors and actuators. However, the lack of secure platforms is also one of its challenges.

Smart cities use DTs to determine the optimal way to maintain critical assets by eliminating guesswork. DT platforms are ideal for leveraging the IoT to boost enterprise services and platforms \cite{Alshammari2021,Stjepandi2021}. Despite its features and benefits, the DT is vulnerable to cyber-attacks due to multiple attack levels and novelty, lack of standardisation and security requirements, and several reasons \cite{BotnSanabria2022,Qian2022}. There are several cyber-attacks in the DT ecosystem, and the type of attack will depend on the cybercriminal's goals. Our study has addressed some challenges that cyber-security DT presents with AI.

\subsubsection{DT in Smart Health}
Using DTs in healthcare can have many social-ethical benefits, including prevention and treatment of disease, cost reduction, patient autonomy, patient freedom, and equal treatment \cite{Alazab2022b}. On the other hand, the social-ethical risks include privacy and property rights, disruption of existing social structures, inequality, and injustice. The confidentiality of patients is a social-ethical issue that is present in personalised medicine. Considering this, it is unclear how DTs will affect the state of these issues and whether they will improve or exacerbate them. The use of DTs requires policies that avoid social and ethical issues and protect individual rights. In addition to protecting data privacy, these policies should protect personal biological information. Data needed for a DT must be high quality, constant, and come from an uninterrupted data stream. Incomplete, inconsistent, and low-quality data can adversely affect performance \cite{Liu.Y.K.2022, Wang.Y.2022}. Cybercriminals may use this information to conduct their activities and launch widespread attacks against these systems \cite{Abounassar2021}. Cybercriminals may infiltrate these systems by compromising confidential, private, and health-related laws. The next are some of the most widespread challenges associated with the adoption of DTs by the healthcare industry \cite{Ho2021}:
 \begin{itemize}
     \item Data Quality: High-quality data are required for a virtual twin to be statistically indistinguishable from its physical twin.
     \item Privacy:  The information must be protected from hackers through data security.
     \item Ethical Concerns: A DT can worsen racial bias, which can cause inequalities in healthcare if a group misrepresents data.
     \item Trust/Fear in AI: Fear of being replaced may prevent healthcare professionals from trusting DT modelling. Healthcare professionals should be supported rather than replaced by DTs.
 \end{itemize}

Additionally, there are concerns regarding the over-collection of data. A regulatory framework should be established to prevent service providers from collecting data irrelevant to the service that the DT service is intended to provide. By collecting more personal information, there is a greater risk of hacking attacks. Detecting hacking and preventing cyber-attacks requires current security measures that are monitored \cite{Zhang.J.2021}. DTs must be reliable to ensure data quality and accuracy, and users should not be able to tamper with them.

To put it another way, users who have their data collected must be unable to exploit the device to compromise its quality and accuracy. Service providers should be able to make sound judgments based on data from the DT if the information collection is stream less, without data corruption \cite{Jimenez2019}. In addition to digital obsolescence, DTs are also subject to the threat of digital obsolescence if the developers do not provide service maintenance following system updates. Users must understand what they consent to when sharing information on DT. Cyberattacks are another potential threat, as DTs can store sensitive information \cite{Pirbhulal2022}. The results of algorithms can be unexpectedly discriminatory. The algorithms used to create models based on the information in DTs may be oblivious to socio-environmental determinants, such as air pollution, water pollution, and inadequate education, all of which can cause health problems. Another challenge when using DTs for personalised health care is the risk of overdiagnosis. Personalised healthcare aims to prevent disease at an early stage. The disadvantage of preventative action occurring too early is that it may lead to overdiagnosis and overtreatment. For enhancing cybersecurity in DT technology for healthcare applications, there are three key elements to be considered (Table \ref{tab:heathcareDT}):

\begin{table}
\scriptsize
\centering
\caption{Reduce the risk of cyber-attacks using DT in healthcare}
\label{tab:heathcareDT}
\resizebox{\linewidth}{!}{%
\begin{tabular}{>{\hspace{0pt}}m{0.27\linewidth}>{\hspace{0pt}}m{0.025\linewidth}>{\hspace{0pt}}m{0.27\linewidth}>{\hspace{0pt}}m{0.025\linewidth}>{\hspace{0pt}}m{0.37\linewidth}} 
\toprule
Attacks &     & Risk Assessment &     & Risk Mitigation \\
\midrule
This stage will demonstrate how attackers may  be able to gain access to the system. With DT, healthcare systems can be  analysed asset-by-asset by providing a continuous overview of  vulnerabilities, threat landscapes, and attack spaces. It is possible to monitor  the chances of attackers entering the system at every step and assess what  kind of security threats they may pose. & $\rightarrow$ & The  organization will be able to gain a deeper understanding of the healthcare  system and its vulnerabilities, risks, and threats because of learning from  the first step. It is also possible to utilize the knowledge acquired from DT  for risk assessment and management at each level at the individual or  operational levels in the future. & $\rightarrow$ & To prevent these risks in IoT-based  healthcare, testing on DT can be performed after identifying the risks and  vulnerabilities associated with each process. The use of DT allows smart  applications to minimize risks throughout each process. Cyberattacks and  risks associated with IoT-based healthcare systems could have devastating  consequences if they are not addressed. As well as compromising data  confidentiality and integrity, cyber-attacks on healthcare services are also  dangerous. Healthcare providers must identify vulnerable processes. CDT can  help improve cybersecurity and combat intruders.  \\
\bottomrule
\end{tabular}
}
\end{table}

\subsubsection{DT in Smart Cars}
The DTs in the truck industry are somehow more advanced than other industries because the increase in the safety of crossing and crossing on the streets and roads depends on the rise of control systems, sensors and the like. Various standards and procedures have recently been published for this purpose in Europe and America \cite{Chen2022,Ozkaya2022}. Most food manufacturing companies must comply with these standards to receive sales standards in Europe and America. Some of these standards are known as DT infrastructure. Every ECU system has a close connection between the hardware, sensors and components of the car with the virtual model of the vehicle in their software, applications and systems \cite{Guo2022B}.

ITS (Intelligent Traffic Systems) have evolved into cooperative or connected vehicles in the past few years. These infrastructures will accelerate the adoption of self-driving cars through data communication between vehicles-to-vehicles (V2V), vehicles-to-infrastructures (V2I/I2V) and other entities on the road \cite{Lv2022b,Rudskoy2021}. Cyber-physical systems have been developed in the last decade, including advanced sensors, subsystems, and intelligent driving assistance applications, which provide unmanned aircraft and vehicles with autonomous decision-making capabilities. Self-driving cars are seen as a serious threat as they must conform to their sensors' structure and degree of complexity and operational applications to be considered autonomous. A thorough analysis of threats and attacks against autonomous vehicles and ITSs, as well as their associated countermeasures, is critical for mitigating these threats. The detection mechanisms of potential attacks on VANETs, ITS, and autonomous vehicles have been the subject of recent reviews \cite{Wu2022}. However, as far as we know, they have not covered actual attacks that have already occurred on self-driving cars. For example, the article \cite{Chowdhury2020} describes the research conducted to identify security gaps and lacks and investigate attacks against self-driving cars. Through these efforts and reports, manufacturers and governments can update their strategies for detecting and countering cyberattacks.

In Europe, the ETSI ITS standard is currently in use. As part of this standard, security has been added as one of the communication layers. The first wave of ITS technology was developed to improve road safety, traffic quality, and road health. Due to its reliance on wireless communication, it may suffer from various hazards that may disrupt its performance and result in serious incidents \cite{Liu.J.2022,Yoshizawa2023}. Sure, the most significant elements of a smart and self-driving car and its attacks and threats are shown in Table \ref{tab:DTcars}.

\begin{table}[htb!]
\scriptsize
\centering
\caption{Attacks on DT cars}
\label{tab:DTcars}
\begin{tabularx}{\textwidth}{XXXX}
\toprule
AV  component        & Threats                                               & Impacts                                                              & Possibility of hazards                                                   \\
\midrule
Infrastructure signs & Change/ add/ remove road sign (e.g., speed limit)     & False or no reaction                                                 & Traffic disturbance, collision, and congestion                           \\
\midrule
GPS                  & Jamming and Spoofing                                  & Inaccurate location info and wrong manoeuvre~                        & Traffic disturbance and crash hazard                                     \\
\midrule
Lidar                & Jamming and smart material (absorbent and reflective) & False detection and degraded Lidar performance                       & Loss of situation awareness and traffic disturbance                      \\
\midrule
Radar/Camera         & Creating blind spots and presenting a false image     & False reaction                                                       & Driver disturbance                                                       \\
\midrule
In-Vehicle device    & Malware and ECU attack                                & Depends on malware capability                                        & Traffic disturbance, disabling vehicle automation and  accident service  \\
\midrule
Acoustic sensors     & Interference and fake sound                           & False Positive/negative obstacle detection and sensor  malfunction~  & Traffic disturbance and low/high speed crash                             \\
\midrule
In-Vehicle sensors   & Eavesdropping and malware                             & Privacy leak, reverse engineering, and false message  generation~    & Serious traffic congestions and driver / traffic  disturbance~           \\
\midrule
Infrastructure (RSU) & Denial of Service and fake WSA (RSA, SPAT)            & Wrong notify to driver, wrong detection, and no  information for ITS & Traffic disturbance, safety issues and critical incident                 \\
\bottomrule
\end{tabularx}
\end{table}

\subsection{The potential uses of DTs in cybersecurity  }
Through DTs, security teams may gain an advantage over sophisticated threat actors and reduce the risks associated with cyber-physical systems in manufacturing, the Internet of Things, and smart consumer devices. DTs can be used in three exciting cybersecurity applications. In cybersecurity, these use cases represent just a sample of what is possible with DTs \cite{Ghimire2022,Sancho2020}. However, while DTs may enable improved detection of anomalous behaviour, the only way to thwart attacks is through accelerated detection and response.  
\begin{itemize}
    \item Secure Design: DTs allow security professionals to simulate a wide range of cyberattacks on physical systems to see how they react when under attack, whether a cyber-physical system within a smart grid, a self-driving car, or an IoT blood pressure monitor. Using these simulated attacks, the design of these devices can be enhanced before they leave the plant floor. To design more robust systems with greater fault tolerance, it is necessary to analyse how the system responds to different cyberattacks. Additionally, DTs enhance the security of a system's design by reducing the attack surface. Leaving aside attack simulations, a detailed analysis of the system's architecture, communications protocols, and traffic flows during regular operations can identify possible weak spots that malicious outsiders can exploit. Removing unnecessary services from the design may be necessary to reduce the system's attack surface \cite{Nunez2020}.
    \item A safer approach to penetration testing: Conducting penetration tests in ICS/OT environments is a valuable but risky activity. It is important to remember that pen tests on live production systems may result in intolerable downtime. Specific paths, techniques, or tools are often not used during these tests due to the risk they pose to availability. As a result, hackers do not care if they take down a crucial operational system to accomplish their real-world objectives. Cyberattacks on OT/ICS environments may well have the downtime of one or more techniques as their primary goal. With DTs, it is possible to perform comprehensive pen tests on the virtual representations of systems without affecting the live system. The dual benefit of this approach is that it addresses more security risks and ensures no downtime.
    \item More intelligent detection of intrusions: DTs can be used to detect intrusions in OT environments, one of the most exciting technologies. Because cyberattacks against these environments are becoming increasingly common due to increasing interconnection, information and control systems (ICS) and distributed control systems (DCS) require intrusion detection so that malicious activities or corporate security policy violations can be precisely monitored. The use of DTs for intrusion detection is outlined in an interesting research paper published in 2020. Due to the ability of DTs to mirror the performance and state of the physical systems upon which they are based, intrusion detection algorithms can monitor and detect attacks rapidly without affecting or interfering with production systems because they can be fed real-time data.  
\end{itemize}

\section{Security and Threats in DT}
This section reviews the DT and IoT environment studies and explores their security issues and recent approaches, including AI, like all innovations, structural limitations, infrastructure, and social acceptance limit DT design, application, and use. A physical cyber system's biggest challenge is creating a digital interface that supports capabilities such as interoperability, trust, stability, reliability, and predictability. Nevertheless, from a technical point of view, data security and human performance, data quality improvement, latency, real-time simulation, large-scale data fusion and aggregation, intelligent data analysis and analysis, forecasting capacity, transparency, and generalisation of technologies in various fields of application are the most critical challenges of DTs.

\subsection{Security Challenges in DT/IoT}
  As the digital revolution spreads, many personal and commercial devices become "smart." In the DT networks, security and privacy may fail. The dynamic nature of IoT connectivity presents a set of security challenges. Here are some examples \cite{Holmes2021}:
  \begin{itemize}
      \item Network scale: In the DT/IoT, billions of smart devices are interconnected in demand and logic, combined with the sheer volume, speed, and structure of real-world data.
      \item Heterogeneity: The DT/IoT intends to connect many heterogeneous devices to implement advanced applications to improve the quality of human life. As a result, IoT devices come in various shapes and sizes, resulting in diverse hardware and software designs. The local policy area also adds to the heterogeneity.
      \item Connection: The DT/IoT links devices and the information they send and receive. Therefore, DT/IoT networks must be available anywhere, anytime, and be able to communicate with other entities under predetermined standards and protocols.
      \item Mobility and dynamism: Network reconfiguration must be dynamic and adaptable since IoT devices are constantly added and removed.
      \item Vulnerability: DT systems are vulnerable to various attacks, such as cookie theft, cross-site scripting, structured query language injection, session hijacking, and even distributed denial of service.
  \end{itemize}

\subsection{Attacks on DT/CPS}
 Over the past years, numerous attacks have occurred against critical infrastructure worldwide, some of which are directly related to DT/CPS, such as Stuxnet 2009, Aramco 2012, Tridium Niagara Framework Attack 2012, Fukushima Daiichi Nuclear Disaster 2012, Godzilla Attack! Turn Back! 2013, German Steel Mill Cyber Attack 2014, Kemuri Water Company Attack 2016, Ukrainian Power Grid Attack 2016, TRITON 2017, Cryptocurrency Malware Attack on SCADA 2018, Norsk Hydro Ransomware Attack 2019, Riviera Beach Ransomware Attack 2019, and Florida Water Treatment Poisoning Attack 2021 \cite{Kayan2022}. Understanding the adversary's tactics is critical to understanding the anatomy of a cyberattack (Table \ref{tab:attackersDT}). Various attacks on DTs are discussed in the following section.
 
\begin{table}[htb!]
\scriptsize
\centering
\caption{A perspective from an attacker on DTs}
\label{tab:attackersDT}
\begin{tabularx}{\textwidth}{p{0.25\textwidth}X}
\toprule
Attack                          & Actuation mode                                                                                                                                                                                                                                                       \\
\midrule
Managing the product  lifecycle & The manipulation of benign behaviour of digital twins to  steer the CPS into an insecure state. Utilise the digital thread to link data  throughout the lifecycle of a product.                                                                                       \\
\midrule
Mode of replication             & Directly update the state of physical devices by  replicating the virtual behaviour of their digital twins.                                                                                                                                                           \\
\midrule
Mode of Simulation              & Re-run test simulations to learn about  system behaviour. The manipulation of simulation parameters or system  specifications' data during security tests may involve exploiting digital  twins' specification-based or machine learning-based process knowledge.     \\
\midrule
The design phases               & Digital twins can be utilised for process knowledge  based on specifications or machine learning.                                                                                                                                                                     \\
\midrule
Phase of decommissioning        & Preserve knowledge about the system's  life for re-use if the digital twin is improperly disposed of. To gain access  to archived digital twins' data, exploit security breaches such as  unauthorised access.                                                        \\
\midrule
Movement in the  laterally      & Obtain control over assets of high value, such as design  artefacts. The new values should be generated randomly to manipulate sensor  readings or simulation parameters at random intervals without causing  significant deviations from the actual process values.  \\
\bottomrule
\end{tabularx}
\end{table}

\subsubsection{Reconnaissance-based attacks}
In a system or platform, identification serves the purpose of collecting information. Following the collection and scanning of ports and I/Os, this information is scanned for security holes and zero-day vulnerabilities \cite{Kulik2022,Pavlov2022}. The Triton malware targeted a petrochemical facility in Saudi Arabia (2012), a successful example of these IT/OT infrastructure attacks. It should be noted, however, that cybercriminals are not necessarily aiming at the core of the system from the beginning; instead, they may attempt to achieve their goals by identifying typical networks and industrial environments. Virtualizers, sandboxes, and similar platforms may also be executed concurrently or asynchronously. As AI tools have become available to cyber criminals, the risks in this field have grown more severe as they can perform network scanning processes without interruption 24 hours a day, seven days a week \cite{Potgantwar2022}. The Stuxnet malware (2009), which Israel introduced into Iran's nuclear energy infrastructure recently, was considered one of the best detection attacks of the past few years. This attack aimed to interfere with the rotation speed of centrifuges after destroying technologies and equipment in the uranium enrichment complex, thereby delaying Iran's nuclear program for several years. Cybercriminals may be able to mount extensive, complex, and effective attacks on DTs \cite{Shi2022}.
\subsubsection{Simulation and CPS }
DTs are complete representations of CPSs. Due to this, the greater the degree of similarity and the greater the connection between CPS and DT, it provides countless benefits for users. It is a cost-effective means of testing various models and increasing the final performance of the simulated version of the system. With DT, copying a CPS and then trying the copied version is unnecessary. In addition to these advantages, DT can also increase the likelihood of an attack succeeding. The cybercriminal records inputs, outputs, and system events in different iterations of the DT in real-time as well as the entire operation of the DT \cite{Kayan2022,Mora2022}. Once the system has returned to a state of process repetition, the attackers simulate its behaviour. For the attacker to effectively utilise the copies of the said information, he must also be actively involved during the repetition process to be aware of the details and synchronisation problems in time-sensitive processes and communicate with the physical entity. As a rule, most of the CPSs are not connected to the simulator version in real-time, but the users' settings and critical parameters entered the system will be the attackers' targets \cite{Hussaini2022}. This situation presents an opportunity to apply AI and learning algorithms since trial and error on the simulator's input data and infinite repetition are very easy for these algorithms. A malicious attacker tests the simulation version of sensitive systems with various input data and attempts to achieve his malicious objectives by rerunning the simulation. Attackers can identify security bugs by reversing settings and configurations and performing security tests.
\subsubsection{Targeting Physical system}
Security in DT's platforms is generally overshadowed by the complexity and importance of implementing details in a DT \cite{Mora2022,Kayan2022,Kulik2022}. Cybercriminals can infiltrate the whole system using the infrastructure used in DTs, which are generally software and middleware platforms. Due to the variety of DT development and security teams' tools, most of these attacks are covert. Through bugs in the system infrastructure, APT has been able to resist software updates over the last few years. In Stuxnet, for instance, the primary purpose of the malware was to record and alter signal transmissions from and to programmable logic controllers, commonly called PLCs. Attributable to the confidential nature of this attack, security experts could not prevent damage to the target hardware, which directly impacted the performance of CPS. AI and ML algorithms in this field can continuously monitor system inputs and outputs during cybercriminal activities. Patterns can be learned, imitated, and changed according to cybercriminals' desired actions. Over the next few years, industrial espionage and cyber warfare will be conducted using this procedure.
\subsubsection{Targeting the DT}
Cybercriminals destroy and lose logical and physical assets by targeting a DT, affecting DT applications \cite{DeBenedictis2022,Kayan2022}. DTs are equivalent to CPS assets since a real system's hardware, PLC codes, and hardware control software entities. Thus, an attack on DT can directly affect the performance of PLCs, controllers, and robots and the nature of the system as a whole. On the one hand, security experts can detect bugs and intrusions using DT. This issue is a double-edged sword; this approach is possible according to the security and safety laws and by analysing the relationship between dynamic variables and recorded data during the time that DT is used. By detecting deviations from the policy defined or learned in a platform, experts in DT systems can provide these findings to security analysts. In contrast, cybercriminals can disrupt the efficiency of DT by detecting system trends and correlating data and variables. Even this action can be performed to prevent the detection of misconduct definitions (based on knowledge or behaviour) and report the details of the attack or destruction. Since the attacker uses the system's assets to carry out the attack strategy (that is, he does not use any software, illegal tools, or malicious tools), no intrusion can be detected. Consequently, detecting long-term deviations in the system is complex and even impossible. If the cybercriminal uses learning tools and AI in this attack, it will be a disaster. As an example, the creation of noise in the system data is not detectable over the long term. Still, it is significant enough to overshadow the platform's performance in DT and, consequently, in the real world of CPS. However, in some CPSs, prototypes of system assets are designed and executed based on simulation during the engineering phases; in other words, a CPS relies on input and output data in DT from conception to completion. DT must, therefore, be able to reproduce the performance of CPS with complete fidelity to details to protect against damages and attacks. In addition, CPS should evolve along with DTs and be compatible with them to the maximum extent possible. Therefore, cybercriminals must continually update their attack techniques.
\subsubsection{CPS/DT Hybrid targeting}
The question is whether it is more dangerous to attack DT and CPS separately or to combine both attacks at the same time? Undoubtedly, each penetration affects both sides, directly or indirectly. Any attack on any CPS/DT ecosystem component will have varying implications depending on its degree of connectivity \cite{DeBenedictis2022,Suhail2022}. Cybercriminals are more likely to target them as their connections become stronger and more real-time. Of course, it should be remembered that cyber warfare and industrial espionage attacks may have a very long reaction time. Preparations for an attack may have been ongoing for several years and focused on preparing for attacks at the appropriate time. Due to this, there do not appear to be any specific countermeasures against these silent attacks, and in some cases, these attacks may only affect DTs, not the CPS. Still, they will overshadow the politics of CPS design in the future \cite{Grasselli2022,Kayan2022}. As a result of most attacks carried out in recent years, the attacker hides on both sides of the ecosystem and in the organisational traffic of the network to increase the complexity of detection for security experts. It is evident that the work has become more complex due to the use of AI, and the data poisoned by AI can replace the input data. Consequently, this plan combines this ecosystem's most destructive and severe attacks.
\subsubsection{Lifecycle and DTs}
DTs go through some phases during their lifetime. The first phase, engineering, uses the CPS model to create knowledge about system processes. Generally, its digital model is developed after preparing a CPS. The project's next phase involves implementing the models, algorithms, and methods for receiving, processing, storing and analysing the system input data from the CPS to the DT. At this stage, the coordination of security models is usually the next priority due to the importance of implementing and modelling a DT from CPS \cite{Kayan2022}. Models created by DT tend to acquire less knowledge than models created by ML because they mimic system behaviour. However, AI models are vulnerable to attacks (data part, model, and algorithm can all be poisoned). Cybercriminals can easily exploit artificial AI if they are too transparent, on the other hand. Cybercriminals can use live data to find CPS execution patterns and act more cautiously when attacking CPSs. However, digital assets occasionally become incomplete due to obsolete or replaced hardware/software in CPS, making it more difficult for cybercriminals to access them. Even so, reviewing system knowledge from previous models is possible because system experts also try to preserve prior knowledge that can be considered in future models. Despite the intentions of experts and system designers, cybercriminals are only interested in exploiting data, models, and algorithms that have been misused. Without a doubt, DTs are attractive targets for these attackers, regardless of cybersecurity concerns.

\subsection{DT's security goals and threats}
In addition to other classifications for DTs, we have used a triple variety to understand the issue better. The DT's platform is divided into three main parts in this category. 
\begin{itemize}
    \item Data collection or sensor side: Infrastructure-level security works on detecting inconsistencies in the physical system and equipment, such as sensors, cameras, and launchers. Most of these pieces of equipment can fall under different cyber terms due to brand incompatibilities.
    \item Network: Communication infrastructure or network part: At this level of security, we mainly focus on anomaly detection, which does not correspond to possible behaviours such as identifying network events, including attacks, accidental access, and data usage. Researchers in this field exploit the unsolved unsupervised learning ability of autoencoders.
    \item Applications: DT platforms and IoT applications are being monitored for malware and botnets at the software level.
\end{itemize}

Cybersecurity literature does not provide a consensus regarding the essential security goals in DTs and IoT infrastructure. Several terms and definitions overlap, e.g., authentication can sometimes be used for identification since both are necessary for each other. There are different definitions of security goals, so this paper is not elaborated on fully. Table \ref{tab:goalsDT} outlines the DT and IoT security goals based on existing literature \cite{Homaei2022,AlTurjman2019,Paul2022}.

\begin{table}[htb!]
\scriptsize
\centering
\caption{Security Goals in DTs and IoT}
\label{tab:goalsDT}
\begin{tabularx}{0.63\textwidth}{p{0.25\textwidth}ccc}
\toprule
Layer           & Sensing   & Network   & Application  \\
\cmidrule{1-1}
Security goals  &           &           &              \\
\midrule
Authorisation   & \checkmark & \checkmark & \checkmark    \\
\midrule
Authentication  & \checkmark & \checkmark & \checkmark    \\
\midrule
Availability    & \checkmark & \checkmark & \checkmark    \\
\midrule
Identification  & \checkmark & \checkmark & \checkmark    \\
\midrule
Integrity       & -         & \checkmark & \checkmark    \\
\midrule
Freshness       & \checkmark & \checkmark & -            \\
\midrule
Confidentiality & \checkmark & \checkmark & \checkmark    \\
\midrule
Privacy         & -         & \checkmark & \checkmark    \\
\midrule
Non-repudiation & \checkmark & \checkmark & \checkmark    \\
\bottomrule
\end{tabularx}
\end{table}

\begin{figure}[h!]
\centering
  \includegraphics[scale=0.33]{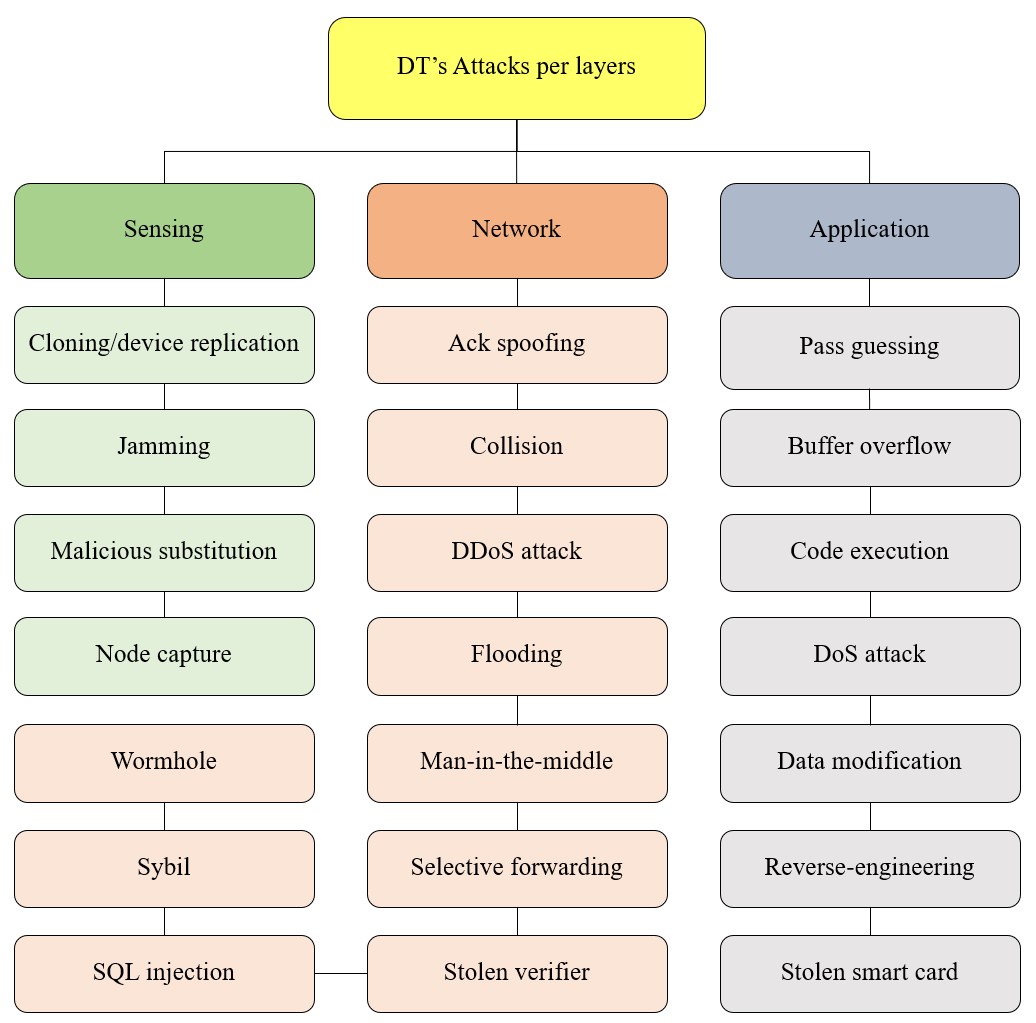}
  \caption{Famous Attacks on DTs and IoT Platformsr}
  \label{fig:7}
\end{figure}

Cybersecurity threats are vast and have various countermeasures to mitigate risks. The DT platform can detect cyber threats such as sensor attacks, spoof-node attacks, hardware manipulation attacks, energy manipulation attacks, sniffing, DDoS, sensitive data leakage, and fault tolerance \cite{Herwig2021,Holmes2021,OlivaresRojas2022}(Fig. \ref{fig:7}).

\section{AI in Cybersecurity}
AI has already penetrated many industries, including cybersecurity. Most businesses believe that they would be unable to detect and withstand attacks without the assistance of AI. According to the Capgemini survey, over 80\% of telecommunication companies use AI for security purposes, and 75\% of banking executives have adopted AI technology. A similar trend can be observed across all major industries. In addition to taking over day-to-day security issues, AI has shown that it can solve complex problems. Protection and hacking \cite{Abdullahi2022,Ansari2022a}. Here is a look at how both sides use AI and what corporations can learn from examining AI applications. The increasing complexity of IT threats is the primary motivation for adopting AI for cybersecurity \cite{Rani2022}. Previously, businesses had to deal with "typical threats" such as Zeus trojans, but new species emerged as we learned how to deal with those threats. Recent threats include Ryuk ransomware, smart botnets, and Triickbot, a newly evolved trojan.

It should be noted. However, evolved threats are not even one of the highlights of risks \cite{Ansari2022b,Arpita2022}. Cybercriminals are equally enthusiastic about utilising AI and ML to their advantage \cite{Kumar2022}. Smart protection methods are necessary to counter equally smart threats already on the horizon. The use of AI in cybercrime is included in some respects \cite{Bonfanti2022}. Even though AI-based attacks have remained rare in the last years, there have been discussions about developing powerful AI threats. Building the infrastructure to maintain such a threat is necessary to create an AI virus. Viruses are small tools with a limited purpose that do not require AI, ML, DT, facial recognition software, or extensive data analysis \cite{Hailu2022,Tsareva2022,Waqas2022}. For the machine to learn, it uses techniques adapted from human learning techniques. These techniques generally work in three stages (Table \ref{tab:MLstepsDT}). 

\begin{figure}[h!]
\centering
  \includegraphics[scale=0.37]{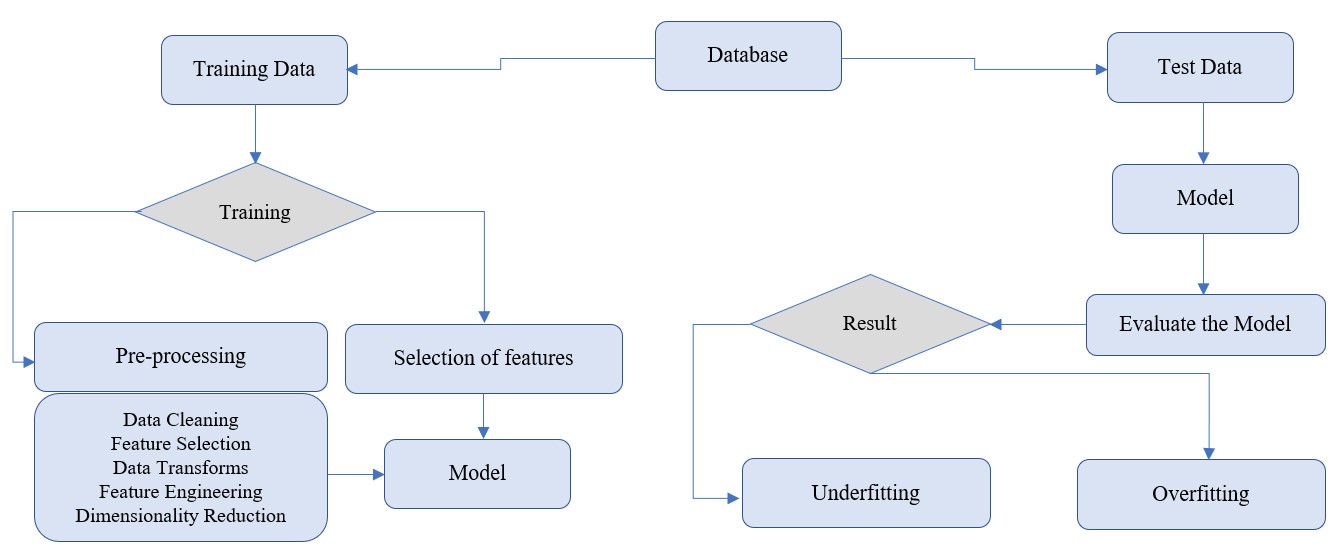}
  \caption{ML and DL models are trained}
  \label{fig:8}
\end{figure}

\begin{table}
\scriptsize
\centering
\caption{Steps in ML and DL in Cybersecurity}
\label{tab:MLstepsDT}
\resizebox{\linewidth}{!}{%
\begin{tabular}{>{\hspace{0pt}}m{0.27\linewidth}>{\hspace{0pt}}m{0.025\linewidth}>{\hspace{0pt}}m{0.27\linewidth}>{\hspace{0pt}}m{0.025\linewidth}>{\hspace{0pt}}m{0.37\linewidth}} 
\toprule
Pre-processing & & Training & & Detection  \\
\midrule
The purpose of this step is to structure the data properly to prepare them for the training phase. & $\rightarrow$ & To build a data model,  useful information has to be extracted from the data prepared before pre-processing. & $\rightarrow$ & Based on the model developed in the previous phase, the supervised traffic data will be used as input to the model, which will compare them with the previous model. Therefore, it can be divided into two classes, normal and abnormal.  \\
\bottomrule
\end{tabular}
}
\end{table}

 Moreover, the attackers exploit these machine-learning techniques to overcome the security system's lines of defence. ML and DL are the best solutions to the security of the line of defence, proving conventional computer security methods are no longer effective. In cybersecurity, an IDS classifies the collected traffic into two main categories (normal and abnormal). Organising it as a different type of attack can go a long way. In this regard, supervised learning is used when a precise outcome is required. In ML and DL, there are supervised/unsupervised and semi-supervised methods, each of which can serve as a tool for cybersecurity. Each has its advantages and disadvantages.
 
The developed models of supervision aim to solve two types of problems: 
\begin{itemize}
    \item Regression problems: attempting to predict the value of an infinitude of continuous variables.
    \item Classification problems: They involve classifying discrete objects with only a limited number of values, as their name suggests
\end{itemize}
	
Fig. \ref{fig:8} is a general demonstration of how learning occurs.

\subsection{Explainable AI for Cybersecurity}
Cybercriminals employ evolving attack patterns to evade detection. A traditional intrusion detection system (NIDS/HIDS), particularly one based on signatures, requires a continuous update. However, growing network data and standard ML tools are lacking as we move towards distributed and real-time processing. Integrating hybrid ML algorithms into big data processing will increase the percentage of accurate results and reveal the hidden knowledge within the data. Compared to manual or statistically-based approaches, ML and DL techniques can reduce the false recognition rate of traditional NIDS \cite{Kaissar2022,Umer2022}. By reducing False Negatives, the detection rate will be improved. Therefore, AI techniques, particularly ML and DL, have been actively studied to assist the IDS in resisting malicious attacks by assuming that the pattern of attack packets differs from the pattern of regular packets. This technique is based on a set of features that identify the state of an object, which must be selected carefully and accurately to avoid inaccuracy and waste of time. Removing unrelated or irrelevant features from the data set is necessary to achieve good data learning with accurate and efficient results. A process such as this is referred to as a feature reduction or a dimensionality reduction. A feature selection process is necessary to select and generate only the most essential features. Some recent AI methods for solving intrusion detection problems include Artificial Neural Networks (ANN), Decision Trees (DT), Evolutionary Algorithms (EA), Genetic Algorithms (GA), Particle Swarm Optimization (PSO), Simulated Annealing (SA), Rule-based Data Mining, and Swarm Intelligence \cite{Anjum2021,BaraaIFarhan2022}.

Different NIDS and HIDS have been deployed using ML/DL techniques in the literature \cite{Zhang2022}. Since each one treats attacks in a particular architecture using different datasets and various ML and DL algorithms, comparing them will not be easy. Like \cite{Capuano2022,Patil2022}, FNN-LSO is proposed for building an advanced detection system and improving the performance of IDS based on Feed-forward Neural Networks (FNN) combined with Locust Swarm Optimization (LSO). Based on their findings, LSO is recommended only for datasets with a high number of features and large datasets. The LSO algorithm is advantageous in large datasets due to the disproportionate number of local optima that render the conventional training algorithm virtually ineffective. Several algorithms have been combined \cite{Aslam2022,Capuano2022} to train a neural network for various real-world datasets using Gravitational Search (GS) and PSO algorithms. Tests of the NSL-KDD dataset were conducted to demonstrate the applicability of the proposed approaches (e.g., error, training time, decision time, overage detection, and accuracy rate) \cite{Hariharan2021,Sharma2022b}.

\subsection{ML/DL in DT cybersecurity}
As wireless connectivity becomes increasingly popular, IoT providers seek the most cost-effective solutions from the outset for their 5G/6G networks \cite{Homaei2021}. Therefore, security must be addressed to combat security threats, including protecting users, network equipment, and data from malicious attacks, unauthorised access, and leakage. A significant role is played by ML and deep learning (DL) in the cyber-security systems of IoT and DT platforms today. As far as security is concerned, DL offers several advantages. Through supervised learning, the system automatically learns the signatures and patterns from experience and generalises them to future intrusions.

Furthermore, it can detect patterns that are different from the usual behaviour of the attackers using unsupervised learning. This definition means that DL/ML can significantly reduce the effort required to redefine the rules to differentiate intrusions. In addition, attackers can use DL to steal and crack users' passwords or other private information. This paper discusses DL/ML-driven network security from three perspectives: infrastructure security, software-level security, and traffic analysis.

\subsection{Attacks via DTs}
Cybercriminals have begun to employ malicious AI to aid attacks, often so that intrusion detection algorithms of the Internet of Things can be thwarted or beneficial AI can work against them (Table \ref{tab:attackMethodDT}). 

\begin{table}[htb!]
\scriptsize
\centering
\caption{Attack method by AI in DT}
\label{tab:attackMethodDT}
\begin{tabularx}{\textwidth}{p{0.25\textwidth}X}
\toprule
Term  & Specifications  \\
\midrule
Cybercriminals can also use Automated Detection of Vulnerabilities & AI as a tool. Attackers often use AI to find and exploit vulnerabilities much faster than developers. Cybercriminals are usually ahead of security experts because they focus on one target \cite{FrancodaSilva2023,Matheu2020}.                                                                                 \\
\midrule
Fuzzing Technique                        & For detecting automated vulnerabilities, fuzzing is proper. Specifically, designed inputs are fed to programs to trigger vulnerabilities and cause the system to crash \cite{Matheu2020}.                                                                                                                                                \\
\midrule
Infiltration Attacks                     & Disruptive inputs to AI-based systems, including noise and algorithm changes by or through physical, cyber, or hybrid methods, expose the digital counterpart of self-driving cars to some of these attacks \cite{Lalouani2022}.                                                                                                            \\
\midrule
False data Injection and Poisoning       & Datasets, models, and algorithms are poisoned by some attacks based on AI. Dataset poisoning directly impacts the understanding of AI and DTs. A method of algorithm poisoning is federated learning, which exploits AI learning algorithms. The legitimate model is replaced with one that has already been poisoned \cite{Lalouani2022}. \\
\bottomrule
\end{tabularx}
\end{table}

\subsection{Countermeasure Via DTs}

Like any technology, AI and ML tools have both positive and negative aspects; in other words, they are cutting both ways \cite{Khan2022,ZixuanZhang2022}. It is critical to understand what the tool's utility is in cybersecurity. Experts use it to improve security, and cyber-attackers gain illegal access to systems. Against the threats and attacks mentioned in the previous section, computer science and security experts have proposed solutions for encryption that can be effective in native digital applications. In summary, we will examine some of these current methods. Blockchain technology is one of the solutions that can be used to increase the reliability and secure the data of a DT \cite{Huo2022,Liu.J.2022}. Since natural and untampered data must feed every DT, it is essential to maintain data confidentiality. Centralised and distributed encryption algorithms allow various industries to distribute and manage data on the blockchain network. It is possible to detect failures and manipulations of codes within the CPS environment via the blockchain network, as data can be tracked and intercepted in any blockchain network. Mechanism and S\&S standards and rules and things like that are applicable. Other advantages of blockchain technology in DT include supporting the engineering phase by providing monitorable and programmable solutions and machine learning patterns. The DTs that benefit from these technologies in various sectors and have multiple execution versions have a high tolerance against attacks and cybercriminals. Another advantage of this technology is that a DT can have several beneficiaries simultaneously, which overcomes the time and place limitations of the beneficiaries. Publishing data in a secure and distributed environment complicates the task for cybercriminals \cite{Suhail2022}. 

Gamification can also be used to increase the security level of the DT. This technique can be accomplished through various security assessments conducted in an isolated environment \cite{Ulmer2022,Vielberth2021}. The procedure described here involves gamification within a non-game environment, such as the game of blue and red teams. In this environment, the agents available for protecting against attacks and the evaluation environment are available for learning. Game environments, or tests, include a management console for assigning roles and resources to security analysts, teams consisting of attackers and defenders for relevant scenarios, and game scenarios with different attack methods. A learning environment is also available in this technology for repeating the DT environment and evaluating and analysing the results. Several tests can be conducted in this field, including security testing, system training, Man-in-the-Middle attacks, penetration, and interference with DT data. Analysing sections describe the steps necessary to learn the system, which should generally be done on a virtual machine to avoid disrupting the DT process. In the end, the security and system development team of the security modules analyses and updates the scenarios and their results. However, despite the mentioned issues, the challenge of the DT's loyalty to CPSs concerning security matters remains an unsolved issue, which makes the efficiency of blockchain and gamification difficult in this field. The following are some popular AI-based methods of these technologies (Table \ref{tab:DefenceDT}).
\clearpage
\begin{table}[htb!]
\tiny
\centering
\caption{Defence method by AI in DT}
\label{tab:DefenceDT}
\begin{tabularx}{\textwidth}{p{0.15\textwidth}XX}
\toprule
Term  & Definition/Efficiency & Limitation/Challenges \\
\midrule
Machine  Learning                    & NB classify data according to the Bayesian theorem, suggesting anomalous activities usually result from multiple events rather than a single attack. Additionally, NB analyses each activity to determine whether it is abnormal after training \cite{Vieira.M.N2022}.                                                                                                                                                                                                  & Zero frequency and the problem of spammers deceiving the NB algorithm led to experts utilising exceptional discoveries such as denying listing to obtain information.                                                                                                                                                                                          \\
\midrule
Federated  Learning Model            & Federated learning offers to exchange data/information security and privacy. The FL model proposes these solutions to prevent and detect manipulation of equipment and user data on a macro scale and to reduce its effectiveness \cite{Alazab2022a}.                                                                                                                                                                                                                  & In the early step of research and development, when the training method and process are still being iterated, federated learning is more expensive than collecting and processing the information centrally.                                                                                                                                                   \\
\midrule
Support  Vector Machines             & Cybersecurity uses SVM's technique to analyse Internet traffic patterns and classify them into HTTP, FTP, and SMTP categories. During penetration testing or network traffic generation, SVMs are often used as training data for attacks that can be simulated \cite{Vitthalrao2020,Adhikari2022}.                                                                                                                                                                             & Due to the learning curve, it is unsuitable for implementing large data sets in  real-world applications. This algorithm is capable of covering a limited number of attacks.                                                                                                                                                                                    \\
\midrule
K-Nearest  Neighbour                 & The K-NN technique is used for intrusion detection and detecting false data injection attacks since it adapts rapidly to new traffic patterns to detect previously unknown attacks\cite{Majeed2021}.                                                                                                                                                                                                                                                                & It requires much memory and is computationally intensive due to storing all the training data.  This algorithm is not used for distributed applications and has a long prediction time, making it unsuitable for real-time applications.  Furthermore, this algorithm is also affected by unrelated features and data sizes.                                   \\
\midrule
Decision  Trees                      & Cybersecurity analyses traffic size, flow, and duration by this method to detect DoS attacks \cite{Vieira.M.N2022}.                                                                                                                                                                                                                                                                                                                                                       & Many correlated and uncertain outcomes can make calculations very challenging when dealing with data classified at different levels. The nodes in a Decision  Graph can be connected by (OR) rather than (AND); whereas the nodes in a  (AND) graph can only be bound by (AND).                                                                                 \\
\midrule
Artificial Neural Networks \par{}~   & ANNs could give rapid response times, which is crucial in systems like traffic management \cite{Zhang.Zh.2021}. Moreover, AI cybersecurity measures often detect or stop attacks in progress rather than preventing them from occurring in the number one place, which is why other preventative security measures should be implemented \cite{Paredes2021}.                                                                                                                         & A neural network faces two mismatch problems: the initial mismatch problem and the inherent mismatch problem. Due to these factors, neural networks are unsuitable for deployment in IT operations management environments.                                                                                                                           \\
\midrule
Explainable  Artificial Intelligence & In contrast to the traditional approach, XAI provides us with an advanced method for making appropriate decisions regarding security background processes. The user receives additional information with the output explaining why a prediction has been completed. This method can generally detect attacks and threats such as Fraud Detection, Intrusion Detection, Spam Detection, Malware Detection, and Traffic Analysis and Identification \cite{Sharma2022b}. & Black box models are these models. It is sometimes difficult to justify the predictions or decisions made due to the opacity of these complex models. Interpretability is one of the most critical factors in cyber security because it promotes trust in the model. Failure to do so may compromise vital information and the organisation's vulnerability.  \\
\bottomrule
\end{tabularx}
\end{table}

\clearpage

\section{Summary}
Due to the increasing growth of technology, their security needs are also increasing. DTs are no exception. There is a need for a new tactic for cybersecurity in the wake of powerful modern viruses and upcoming AI-based viruses \cite{Ansari2022a,Bonfanti2022,Vieira2022}. There is no doubt that organisations must stay one step ahead of threats. The fact that cybercriminals have not actively targeted us with AI-based threats yet does not mean it will not happen soon. For several reasons, businesses must adopt AI in their security systems. Some reasons for the vital need for DT cybersecurity to use AI are as follows. AI and machine learning are essential for identifying and combating threats that find new ways to penetrate personal devices and corporate databases \cite{Shandilya2022}. For example, AI and ML are not limited to attacking known strains of viruses - they can also affect unknown strains. A threat can be checked in the existing database, similar patterns can be detected through smart algorithms, and a counterattack strategy can be generated. Over time, such a system becomes more adept at detecting and dealing with unknown threats, increasing its efficiency and speed. Cybercriminals often disguise malicious code among bundles of meaningless combinations, making it virtually impossible for a typical antivirus to detect dangerous components. A security software based on ML detected OceanLotus, one of the Vietnamese hacking organisations, using this strategy. The use of AI in security has other potential applications, and Table \ref{tab:AIcybersecurityDT} represents some of these potentials.

\begin{table}[htb!]
\scriptsize
\centering
\caption{Potential applications of AI in cybersecurity of DTs}
\label{tab:AIcybersecurityDT}
\begin{tabularx}{\textwidth}{p{0.3\textwidth}X}
\toprule
\textbf{Potential} & \textbf{Advantages}\\
\midrule
Detecting viruses                   & AI can analyse terabytes of data within a brief period, detecting suspicious code fragments in a short amount of time.    \\
\midrule
Creating a virus database           & AI will store, process, and learn from previously detected threats.                                                        \\
\midrule
Anticipating cybercriminals'  moves & AI can analyse existing threats, security news, and trends to forecast future developments.                                \\
\midrule
Optimising functionality            & AI can help businesses improve their software and decrease the likelihood of future attacks by providing smart insights.  \\
\bottomrule
\end{tabularx}
\end{table}

Despite the high potential of AI in providing and increasing the productivity of cybersecurity in DT, these platforms also suffer from risks. In other words, AI has created a threat and an opportunity where the role of researchers and governments is vital, and their role is essential. Some threats of AI in the cybersecurity of DTs are given in able \ref{tab:AIthreatsDT}.
\begin{table}[htb!]
\scriptsize
\centering
\caption{AI threats in cybersecurity of DTs}
\label{tab:AIthreatsDT}
\begin{tabularx}{\textwidth}{p{0.3\textwidth}X}
\toprule
Threat & Definition \\
\midrule
The random mutation of executable code  & Polymorphic viruses are already capable of doing this,  but AI can increase the variety of variables supported.                                                                         \\
\midrule
Operating system adaptability           & The AI-based virus could find an intelligent approach to Kerner-level functions or apply rootkits to avoid being detected.                                                            \\
\midrule
Identifying the antivirus and attacking & AI viruses can detect anti-virus software and develop methods for attacking its code.                                                                                                  \\
\midrule
Detection of social networks            & Viruses can mimic human language and trick users into sending confidential documents, providing access data, or simply perpetrating cyberbullying using conversational programming.  \\
\midrule
Update creation                         & Once an antivirus detects the previous version, it can push a  new version and continue its malicious activities.                                                                       \\
\bottomrule
\end{tabularx}
\end{table}

Viruses in the next generation will likely utilise cyber technology \cite{Aloqaily2022,Kumar2022}. We have not seen many such threats because of financial and technological limitations: viruses are often created by individuals who do not have access to robust infrastructure that would allow them to construct AI \cite{Ansari2022b}. It is expected, however, that the technology will become more accessible over time and that individual hackers can implement it in their attacks. The implementation and security of DTs by AI include several challenges mentioned in \ref{tab:AIchallengesDT}. The active use of AI and ML is not the only challenge organisations and cybersecurity professionals must overcome \cite{Abdullahi2022,Ansari2022b,Vieira2022}. There are also other problems associated with the current approach to security.
\begin{table}[htb!]
\scriptsize
\centering
\caption{AI and cybersecurity challenges in the DT}
\label{tab:AIchallengesDT}
\begin{tabularx}{\textwidth}{p{0.3\textwidth}X}
\toprule
Challenge & Definition \\
\midrule
Distanced infrastructure        & System communications today span continents, sending sensitive data around the globe. Insufficient protection is in place for these transfers, making them easier to hack.                                                                              \\
\midrule
The detection process is manual & Security threats and suspicious patterns are not the focus of human teams 24 hours a day, seven days a week. It is common for systems to go unmonitored most of the time.                                                                                \\
\midrule
Security teams’s reactivity     & It is more common for security experts to focus on responding to threats rather than predicting them.                                                                                                                                                     \\
\midrule
Providing dynamic threats & Hackers use several strategies to conceal their locations, IP addresses, identities, and methods.  In contrast, the cybersecurity field is much more transparent and open to research - data created by businesses are readily available to criminals.  \\
\midrule
Heterogeneity/Scale             & A wide variety of hardware and software designs resulting from  DT/IoT platforms as well as different local policy areas also add to the heterogeneity and difficulties of applying artificial intelligence in cyber security.                           \\
\bottomrule
\end{tabularx}
\end{table}

Regarding technological trends and advancements, AI fits nicely into the larger picture. Cybersecurity challenges are primarily posed by technology. In addition to providing continuous oversight, it can detect distributed threats and learn about new attacks as they occur. Contrary to standard solutions, it can detect and deal with unknown agents and existing and catalogued threats. Here are certain of the main advantages of using AI and ML in security that has made AI a leader in the current global security market (Table \ref{tab:AIadvantagesDT}).

\begin{table}[htb!]
\scriptsize
\centering
\caption{Advantages of using AI in DT cybersecurity}
\label{tab:AIadvantagesDT}
\begin{tabularx}{\textwidth}{p{0.3\textwidth}X}
\toprule
Advantage & Definition \\
\midrule
Fast detection                & The capabilities of AI exceed those of human beings in terms of analysis and monitoring. AI  has a significant advantage over traditional processing methods: it can identify unknown threats and create response strategies from scratch.                                                                                                      \\
\midrule
No human errors               & Handling small tasks daily, AI can assist in making crucial strategic decisions. Without checking their decisions with the smart, data-driven algorithm, businesses risk overlooking an important piece of data or failing to recognise a hidden pattern.                                                                                     \\
\midrule
Quick response                & Attackers can process terabytes of data in seconds with AI tools. Even large corporations can detect threats within seconds with AI and ML solutions.                                                                                                                                                                                           \\
\midrule
Automating routine work       & The security team can devote more time to visionary and strategic objectives with AI solutions. Regarding creativity and strategic insight, human experts are still far superior to their automated counterparts. Hence, it makes sense to free them from easily automated tasks so that they can concentrate on the most critical processes.  \\
\midrule
A smart approach to education & AI can be used to gather information and generalize it about current viral threats, create smart databases, identify risks, and respond accordingly.                                                                                                                                                                                            \\
\bottomrule
\end{tabularx}
\end{table}

Even though AI can solve most cybersecurity challenges, it is not the ultimate solution. AI is currently difficult to implement on a small scale, and the technology requires improvement. According to recent research, the advantages of AI solutions outweigh the disadvantages; however, this does not diminish the fact that AI solutions have several drawbacks compared to human beings (\ref{tab:AIdisadvantagesDT}). 

\begin{table}[htb!]
\scriptsize
\centering
\caption{Disadvantages of using AI in DT cybersecurity}
\label{tab:AIdisadvantagesDT}
\begin{tabularx}{\textwidth}{p{0.3\textwidth}X}
\toprule
Disadvantage  & Definition\\
\midrule
Hackers are also  AI-savvy      & AI security solutions can also be used by hackers.  Compared to cyber criminals, cybersecurity teams have much more open practices. In contrast, hackers are unlikely to benefit from organisations'  progress, while they may be able to reverse their findings and introduce a  more effective threat.                                                             \\
\midrule
Cyberthreats continue to evolve & Integrating artificial intelligence into your business does not automatically make you immune to all threats. Even AI systems must be constantly redesigned, improved, and maintained as viruses and malware improve.                                                                                                                                                 \\
\midrule
Adoption hurdle                 & Unlike traditional antiviruses,  artificial intelligence still requires plenty of human resources and computing power. Instead of spending time and money developing a custom AI  solution, you can install ready software. Despite this, AI is becoming increasingly available, and even small businesses can afford to construct a  neural security network.  \\
\bottomrule
\end{tabularx}
\end{table}

\section{Conclusion}
 The continuous connection of things to the global Internet network and the growing use of communication and monitoring technologies have enabled artificial intelligence and machine learning to be applied in recent years. DT will become a leading player in the monitoring and controlling processes in almost all industries and our daily lives. Several ongoing developments in industry, transportation, medicine, air and space, agriculture, the environment, and many others suggest that every entity in the world will have a digital counterpart. This article examines what, why, and the place of DT in various applications and their challenges. Cybersecurity is one of the most significant challenges for DT in the cyber environment. Because threats to a DT directly affect its CPS and vice versa. As a result of this review article, we answered some essential questions: What are the security challenges, attacks and threats facing DTs? How to secure them? What are the necessary tools to provide security in DT and, consequently, CPS?
 
 In this regard, we examined some applications to note the necessity of dealing with cybersecurity in the initial stages of designing any DT. In the following, we have mentioned artificial intelligence solutions and how to use these innovative tools. If cyber criminals take possession of artificial intelligence tools, we will be able to deal with and secure DT only with similar tools. The wide range of DT applications has made it difficult to investigate all security risks, attacks, threats, and methods to deal with them. Nevertheless, in this article, we tried as much as possible to draw the attention of researchers in cybersecurity, digitalisation, monitoring and computer science to this issue. Apparently, this field will be investigated more in the coming years. In the end, it should be emphasised again that cybersecurity challenges in the field of DTs are worth exploring to advance this concept.

\bibliographystyle{elsarticle-num-names} 
\bibliography{manuscript}

\end{document}